\documentclass[aps,prx,reprint,onecolumn,notitlepage,superscriptaddress]{revtex4-1}
\usepackage{amsmath,ulem}
\usepackage{eucal}
\usepackage[usenames,dvipsnames]{color}
\usepackage{graphicx}
\usepackage{amssymb,amsfonts,amsmath}
\usepackage{bm}
\newcommand{\off}{\mathit{b}}
\newcommand{\thr}{\mathit{thr}}
\newcommand{\hit}{\mathit{hit}}

\begin{document}

\title{Odor Landscapes in Turbulent Environments}
\author{Antonio~Celani}
\affiliation{Institut Pasteur, Research Unit ``Physics of Biological Systems", 75724 Paris Cedex 15, France, and
CNRS UMR 3525,  28 rue du docteur Roux, 75015 Paris, France}
\affiliation{The Abdus Salam International Centre for Theoretical Physics (ICTP), Strada Costiera 11, I-34014 - Trieste, Italy }
\author{Emmanuel Villermaux}
\affiliation{Aix Marseille Universit\'e, CNRS, Centrale Marseille, IRPHE UMR 7342, 13384 Marseille, and Institut Universitaire de France, 75005 Paris, France}
\author{Massimo~Vergassola}
\affiliation{Institut Pasteur, Research Unit ``Physics of Biological Systems", 75724 Paris Cedex 15, France, and 
CNRS UMR 3525,  28 rue du docteur Roux, 75015 Paris, France}
\affiliation{University of California San Diego, Department of Physics, La Jolla, CA 92093 USA}

\begin{abstract} 
The olfactory system of male moths is exquisitely sensitive to pheromones emitted 
by females and transported in the environment by atmospheric turbulence. Moths respond to minute amounts of pheromones and their behavior is sensitive to the fine-scale structure of turbulent plumes where pheromone concentration is detectible. The signal of pheromone whiffs is qualitatively known to be intermittent, yet quantitative characterization 
of its statistical properties is lacking. This challenging fluid dynamics problem is also relevant for entomology, neurobiology and the technological design of olfactory stimulators aimed at reproducing physiological odor signals in well-controlled laboratory conditions. Here, we develop a Lagrangian approach to the transport of pheromones by turbulent flows and exploit it to predict the statistics of odor detection 
during olfactory searches. The theory yields explicit probability distributions for the intensity and the duration of pheromone detections, as well as their spacing in time. Predictions are favorably tested by using numerical simulations, laboratory experiments and field data for the atmospheric surface layer. The resulting signal of odor detections  lends to implementation with state-of-the-art technologies and quantifies the amount and the type of information that male moths can exploit during olfactory searches.

\end{abstract}

\maketitle


\section{Introduction}
Sex pheromones provide arguably the most striking example of long-range communication through specialized airborne messengers \cite{W03}.

Most Lepidoptera are consistently attracted to calling females from distances going as far as several hundred meters, reaching their partners in a few minutes \cite{WP87}. 
This feat is impressive as females broadcast their pheromone message into a noise-ridden transmission medium, the turbulent atmospheric surface layer, and receiver males face the challenge of extracting information about the female's location from a signal that is attenuated, garbled and mixed to other olfactory stimuli (see Fig.~1).

The pheromone communication system is under strong evolutionary pressure. This is particularly evident for adult moths of the family {\it Saturniidae} and {\it Bombycidae} (e.g. the indian luna and the silk moth, respectively), which have a lifespan of a few days as adults.  Subsisting on stored lipids acquired during the larval stage, they largely devote their adulthood to the task of reproduction. The result of natural selection is  an olfactory system exquisitely sensitive to pheromones\,: just a few molecules impinging on the antenna of a male moth are sufficient to alert the insect and trigger a change in its cardiac frequency \cite{A03}; concentrations of few hundred molecules per cubic centimeter elicit specific behavioral responses that prelude flight \cite{BKS65}.

The quality and the time-course of the pheromone signal matter, in addition to its intensity. As for the quality,
the signal is usually a blend of two or more chemical compounds. Species of closely related families often use similar components and discrimination is achieved by different combinations and/or ratios in the mixture. Pheromone
components of sympatric species that emit
similar pheromone blends, often act as behavioral antagonists \cite{BC79} and the discrimination
among different blends is extremely fine \cite{BFC98}. The first-order center for the discrimination is the macroglomerular complex of the antennal lobe, where detections from olfactory receptor neurons are integrated \cite{R12}. As for the time-course of the signal, turbulence strongly distorts the pheromone signal,
leading to wildly intermittent fluctuations of concentration
at large distances from the source. As shown in Fig.~1, the signal features alternating bursts and clean-air periods with a broad spectrum of durations \cite{MEC92}. 

Characterizing the properties of odor detections in turbulent flows is a challenging and fundamental problem in statistical fluid dynamics. Furthermore, 
intermittency generated by the physics of turbulent transport is crucial for eliciting the appropriate biological behavior. Insects exposed to steady, uniform stimuli briefly move upwind, arrest their flight toward the source and begin crosswind casting (the typical response to the loss of olfactory cues). Males temporarily resume upwind flight when the stimulus is increased stepwise, and set into sustained upwind flight when exposed to repeated pulses \cite{K81,WB84,B85}. Hence, the statistics of turbulence-airborne odor stimuli is literally the message sent by female to male moths, it controls their behavior and defines the information that male moths can exploit for their searches \cite{Bossert63,Bossert68,FMLLC02,LFC01}. Therefore, the long-standing problem of characterizing the statistics of odor detections during olfactory searches is essential to understand the neurobiological response of insects \cite{Hopfield91}. Additional motivation for considering the problem stems from laboratory experiments using olfactometers and/or tethering.  
 Experiments in \cite{DCF09,B10} have {\it Drosophilae} tethered to a wire and assay their responses (electrophysiologically and/or behaviorally) to simple odor stimuli, such as pulses of fixed duration, that are most likely not representative of those experienced in the wild. Determining the statistics of physiological stimuli, to reproduce it then in the laboratory, would represent a major progress and significantly impact the design of future experimental assays. 

Here, we address and answer the following questions\,: How intermittent is the distribution of pheromones as a function of the down/cross-wind distance from the source\,? What are the statistical distributions for the intensity and the duration of odor-laden whiffs, and the duration of clean-air pockets\,? What is the dependency on the sensitivity threshold\,? How does turbulence affect the ratio among different components of a blend from emission to reception\,?  Can emissions from multiple sources, with different blend ratios, reach the receiver without being irremediably mixed\,? Results are obtained by developing a theoretical Lagrangian approach that predicts the salient properties of a tracer emitted by a localized source and transported by a turbulent flow. We focus on a continuously emitting source yet methods  generalize to periodic emissions. Predictions are successfully tested by numerical simulations, laboratory and field experimental data. Consequences for the neurobiological responses of insects during olfactory searches and for laboratory protocols of olfactory stimulation are discussed in the conclusions. 

\section{Theoretical framework}

\paragraph*{\bf Definition of the problem.--}
We  consider the emission by a source of linear size $a$ (at the origin ${\bm x}={\bm 0}$) of a chemical substance (or a mixture) at a constant rate of $J$ molecules per unit time. The environment transporting the chemical is a turbulent incompressible flow ${\bm u}({\bm x},t)={\bm U}+{\bm v}({\bm x},t)$. The mean wind is ${\bm U}=(U,0,0)$  while ${\bm v}$ is the turbulent component. The turbulence level $v/U$, that is the ratio between the amplitudes of the turbulent component ${\bm v}$ and of the mean flow ${\bm U}$, is supposed small in the rest of the paper.  We are interested in the time-series of the concentration $c$ at a downwind distance $x$ (much larger than $a$ but still smaller than the correlation length $L$ of the flow) and crosswind distance $y$ from the source (see Fig.~1). The concentration $c({\bm x},t)$ of the chemical  obeys the advection-diffusion equation 
\begin{equation}\label{eq:M1}
\frac{\partial c({\bm x},t)}{\partial t} + {\bm u}({\bm x},t) \cdot {\bm \nabla}  c({\bm x},t) = \kappa \nabla^2 c({\bm x},t) + J h_a({\bm x})\,,
\end{equation}
where $\kappa$ is the molecular diffusivity. The function $h_a({\bm x})$ is the spatial
distribution of the source of size $a$, e.g. a top hat vanishing outside the source ($|{\bm x}|> a$) and
  normalized to unity ($\int h_a({\bm x}) d{\bm x}  = 1$). 

\paragraph*{\bf Quantities of interest.--}
We shall derive below the expressions for the following observables of the concentration field $c$ at a given spatial location (see Fig.~1): ({\it i}) The intermittency coefficient, $\chi$, defined as the fraction of time  the concentration is non-zero. The smaller this number, the longer the insect performing the olfactory search is exposed to clean air; ({\it ii}) The average concentration $C$ taken over periods of time when the signal is non-zero. The value of $C$ determines the typical intensity of concentration in an odor-laden plume and whether or not that level is detectible by the insect, as discussed below; ({\it iii})  The full statistics of the signal intensity, that is  
the probability distribution $p(c)$ of the concentration. Its expression involves $C$ and $\chi$ as fundamental parameters. ({\it iv})  Insects are supposed to detect a signal during those intervals of time when the local concentration exceeds some sensitivity threshold $c_\thr$. We call those periods ``whiffs'', whilst the complementary periods when $c\le c_\thr$ are dubbed ``blanks'', or ``below threshold''. The temporal structure of the signal is thus given by $p(t_w)$, the probability  distribution of the duration $t_w$ of the whiffs, and by $p(t_b)$, the probability distribution of the duration $t_b$ of intervals below threshold, which we obtain below.

\paragraph*{\bf The Lagrangian approach.--}

Lagrangian methods (see \cite{Durbin,Hunt,Pope,FGV01,SS00,Sawford} for introduction and reviews) focus on fluid-parcel trajectories and the statistics of the concentration field is reconstructed from the properties of a suitable ensemble of trajectories.  Lagrangian approaches are alternative to the Eulerian description, where the main focus is the concentration field itself (as, e.g., in the fluctuating plume model \cite{MEC92}). The two descriptions are formally equivalent yet they lend to physical approaches that are quite distinct. The Lagrangian reformulation of  \eqref{eq:M1} is
\begin{equation}\label{eq:M2}
c({\bm x},t) = J \int_{-\infty}^t dt' \int d{\bm x}'\, h_a({\bm x}')\, p_{{\bm v}}({\bm x}',t'|{\bm x},t)\,, 
\end{equation}
where $p_{\bm v} d{\bm x}'$ is the probability that a fluid parcel transported by the flow is around 
${\bm x}'$ at time $t'$, given that it is in ${\bm x}$ at time $t$. The index of $p_{\bm v}$ is meant to stress that the probability is averaged over the molecular noise statistics but no average is taken over the fluctuating turbulent flow ${\bm v}$ (more details can be found in \cite{FGV01}).
Eq.~(\ref{eq:M2}) states that $c({\bm x},t)$ is determined by tracing back in time the trajectories of parcels that end in ${\bm x}$ at time $t$. The ensemble of those trajectories forms a puff whose center of mass recedes upwind and whose size $r(t')$ typically grows as $t' \to -\infty$ (see Fig.~2). Depending on the realizations of ${\bm v}$, two cases can be distinguished\,: (i) the distance between the center of mass of the puff and the source never becomes smaller than the size of the puff. These are pockets of clean air, where the concentration $c({\bm x},t)$ vanishes, as it follows from \eqref{eq:M2}; (ii) otherwise, the concentration $c({\bm x},t)$ is non-vanishing. 

It follows from (\ref{eq:M2}) that the value $c({\bm x},t)$ is proportional to the time of overlap between the puff and the source. The problem thus reduces to characterizing the statistics of the corresponding residence time. 
The turbulent flow that disperses the puff creates convoluted folds of local structures having some directions extended while the others are contracted down to the diffusive scale $\eta$ of the scalar concentration field. The specific nature of those structures is determined by the signs of the Lyapunov exponents of the flow. Here, though, we are 
interested in the statistics of the residence time at the source, the size of which is $a\gg \eta$. Therefore, we physically expect that the small-scale structures of the puff are smoothed out by the integrals appearing in (\ref{eq:M2}), affecting only constant factors that are not essential for the specific quantities discussed here. In particular, if we disregard constants of order unity, a sufficient characterization of the puffs should be provided by 
 the dynamics of their center of mass and their overall size. We shall derive below the consequences of these physical assumptions and compare the resulting predictions to numerical and experimental data.

\paragraph*{\bf Lagrangian properties of the turbulent flow.--}
\label{expos}

We will show shortly that the statistics of odor stimuli for the problem defined above depends on the details 
of the turbulent flow transporting the pheromones via three exponents $\alpha$, $\gamma$ and $\beta$. Power laws are typically observed in turbulent flows as a consequence of  scale-invariance properties of fluid dynamical equations \cite{Uriel}. The exponents that we define below are 
related respectively to the dynamics of single-particle, pair dispersion, and rate-of-growth of the size of a dispersing puff.

(i) The exponent $\alpha$ controls the distance travelled by a single particle at short times $t$ as $(kt)^{1/\alpha}$, with $k$ constant. The crosswind width of the average plume,  outside of which detections are  rare, scales with the downwind distance $x$ as $x^{1/\alpha}$. In most physical cases, the mean wind gives the dominant contribution, so that  $\alpha=1$, $k=U$ and the shape of the average plume is conical. 
However, for one special case discussed below (the Kraichnan flow), single-particle dispersion is dominated by diffusion at short enough times ($\alpha=2$) and the standard $Ut$ behavior holds only at longer times (yet smaller than those needed to reach the source). 

(ii) The exponent $\gamma$ is related to the dispersion of a pair of particles as $(k't)^{1/\gamma}$, where $k'$ is a constant. For the applications  below, the relevant values are $\gamma=2/3$, corresponding to the Richardson-Kolmogorov scaling, $\gamma=2$ for ordinary diffusion and $\gamma=1$ for ballistic separation. 

(iii) Finally, the exponent $\beta$ is defined by the scaling relation for the rate-of-growth 
$ \zeta_{r,t} \equiv d\log r/dt =t^{-1}(k't/r^\gamma)^\beta$

of  a puff of size $r$ at time $t$ after its release. For homogeneous and stationary flow, $\beta=1$ and $\zeta$ depends on the size only. However, if the flow is inhomogeneous, the dependency is more complicated. Namely, in the neutral atmospheric layer the dynamics explicitly depends on the height and the height of particles released close to the ground grows linearly with time. Non-homogeneous effects of the height are then conveniently accounted via the dependency of $\zeta_{r,t}$ on the time $t$ since the release of the puff (we show below that $\beta=2$ in this case). The consistency between the definitions of $\beta$ and $\gamma$ is easy to check\,: 
$d r/dt =r\zeta_r\sim k'^{\beta} t^{\beta-1} r^{1-\beta\gamma}$ and integration of the equation yields $r\sim (k't)^{1/\gamma}$ for any $\beta$.


\section{Results: Theory}

In this section we summarize the theoretical results about intensity and dynamics of the concentration signal. The derivations are detailed in the Appendix A.

\paragraph*{\bf The intensity of the concentration signal.--} 
We first consider statistical objects that quantify the concentration $c$ of the pheromones at a given time. The intermittency factor $\chi$ is defined as the fraction of time that $c$ is non-vanishing; the average of the concentration $c$ over that fraction of time is denoted $C$. The threshold of detection, i.e. the minimum concentration that the receiver is able to sense, is denoted $c_{\thr}$.  Intervals when $c> c_{\thr}$ are ``whiffs'', while ``blanks'' are the complementary regions $c\leq c_{\thr}$ when the signal is either absent or not detectible. The ratio $C/c_{\thr}$ controls whether or not a typical plume is detectible. Using Lagrangian methods, we show in the Appendix A that 
\begin{equation}\label{maineq:0}
\begin{array}{l}
\chi =\mathrm{Prob}(c>0)\sim 
\left(\frac{k' x^{1-\gamma}}{U}\right)^{(3-\alpha)/\gamma}f\left(\frac{U y^\alpha}{k x} \right)\,,
\\
 C =\left\langle c | c>0 \right\rangle \sim 
 \frac{J}{k} \left(\frac{k'x}{U}\right)^{-(3-\alpha)/\gamma}\,,
 \end{array}
\end{equation}
where $f$ is a nondimensional function that decays rapidly for large arguments,
namely exponentially in the applications below.
Eq.~(\ref{maineq:0}) indicates that $\chi$ decreases and $C$ remains constant, as $y$ increases.
Therefore, moving crosswind away from the mean-wind axis, the signal retains its intensity but becomes sparser.
Approaching the source (reducing $x$), the intensity within a whiff grows, 
while the frequency of encounters depends on $\gamma$. 
We show in the Appendix A that the concentration $c$ is inversely proportional to the size of the Lagrangian puff (see Fig.~2) when it hits the source. Intense concentrations are associated to flow configurations which leave the  puff atypically small. Using that the occurrence of those configurations is a rare event that obeys a Poisson statistics, we show then that the tail of the probability distribution $p(c)$ is\,:
\begin{equation}\label{maineq:1}
p(c) \sim \frac{\chi}{C} \left(\frac{c}{C}\right)^{-2+\frac{\beta\gamma}{3-\alpha}} \exp\left[-\left(\frac{c}{C}\right)^{\frac{\beta\gamma}{3-\alpha}}\right] \,.
\end{equation}
for $C \ll c \ll c_0$ where $c_0$ is the concentration at the source.
The moments $\langle c^n\rangle$ are shown (see Eq.~\eqref{eq:M11}) to depend on $C$ and $\chi$ in \eqref{maineq:0} via the relation $\langle c^n \rangle \sim \chi C^n$, consistently with the scaling form \eqref{maineq:1}.

\paragraph*{\bf The duration of the whiffs.--}
Since the behavior of the insects depends on the time-course of the odor stimuli, it is important to characterize the statistics of the whiffs, i.e. time intervals when the concentration is above the threshold $c_{\thr}$ of detection. We predict (see Appendix A) for the distribution of  the duration $t_{w}$ of the whiffs
\begin{equation} \label{maineq:3}
p(t_w) \sim  \tau^{-1} \left( t_w/\tau \right)^{-3/2} g_w\left(t_w\right)\,.
\end{equation} 
The power law $-3/2$ is cutoff by the function $g_w$, constant for small arguments, decaying exponentially with rate $T_w^{-1}$ for $ t_w \gtrsim T_w$ and eventually crossing over to a stretched exponential $\propto -t_w^{\beta}$ for $t_w\simeq x/U$   (the typical time to reach the source).   The  cutoff $T_w$  is determined by two physical mechanisms (see Fig.~2d)\,: (i) the flow changes in time and its new configuration is more effective at dispersing the puff, increasing its size and making the concentration fall below the threshold $c_{\thr}$; (ii) large-scale velocity fluctuations displace the puff away from  the source. The expression for the corresponding cutoffs $T_1$ and $T_2$ is derived in the Appendix A, and $T_w$ is the minimum between the two. The relative importance of the two mechanisms depends on the details of the flow transporting the odors, on the distance to the source and on the threshold $c_{\thr}$, as discussed in the examples below.

The power $-3/2$ in \eqref{maineq:3} originates from the wiggling of the Lagrangian puff in Fig.~2 around the source, 
leading to the alternation of whiffs (overlaps with the source) and blanks (loss of overlap) distributed according  to the properties of a diffusion process.
The parameter $\tau$ is the shortest overlap, i.e. the time to diffuse across a distance $\simeq a$,  the size of the source.
Due to the slow power-law decay $-3/2$ in  \eqref{maineq:3}, \textcolor{black}{the average duration is determined by the cutoff\,: $\langle t_w \rangle \sim (\tau T_w)^{1/2}$.} 

\paragraph*{\bf The duration of the blanks.--} Blanks are time intervals when the concentration is below  the threshold $c_\thr$ and thus no signal is detectible.
For the probability density of their duration $t_\off$, we derive in the Appendix A
\begin{equation}\label{maineq:4}
p(t_\off) \sim \tau^{-1} \left( t_\off/\tau \right)^{-3/2} g_\off(t_\off)\,.
\end{equation}
Here, $g_\off$ is approximately constant for durations shorter than the cutoff $T_\off$ and then decays exponentially with  rate $T_\off^{-1}$.
The identical $-3/2$ power laws in eqs.~\eqref{maineq:3} and \eqref{maineq:4} stem from the short-time diffusion of the Lagrangian puff around the source (see the Appendix A for more details), which symmetrically looses and gains contact with the source. Note that the power laws do not depend on the details of the flow. The temporal structure of whiffs and blanks contains then some information which is independent of environmental variations of the intensity, stratification and other details of the flow transporting the pheromones. 
It follows from \eqref{maineq:4} that \textcolor{black}{ the average duration of the blanks $\langle t_\off \rangle\sim (\tau T_\off)^{1/2}$}, i.e. it is determined by the cutoff of the distribution, as for the whiffs.
Since whiffs and blanks are mutually exclusive, their averages (and thus their cutoffs) are not independent. The exact relation \eqref{eq:Toff} derived in the Appendix A shows indeed that the ratio of the two averages is given by the ratio of the probabilities that the concentration is above or below the threshold of detection. 
Specifically, \eqref{maineq:0} and \eqref{maineq:1} indicate that the value of $C$ and the statistics of $t_w$  do not depend on the crosswind distance, i.e. the whiffs do not change in their intensity and duration while moving crosswind. Their frequency does change, though, which reflects in the intermittency factor $\chi$ in \eqref{maineq:0} and affects the statistics of blanks. In particular, the cutoff $T_{\off}$ will grow while moving crosswind according to \eqref{eq:Toff}.

\paragraph*{\bf Clumps of whiffs.--} The visual counterpart of the broad distribution \eqref{maineq:3} for the  whiffs is their aggregation in clumps, as in Fig.~1.
The short-time diffusion of the Lagrangian puff discussed at point (iii) in Section~\ref{durwhiffs} implies that on/off times within a clump have the same statistics as the time intervals spent above/below zero by a random walk with time step $\tau$. 
As a result, the total number of whiffs in a clump of size $T_w$ is typically $\sqrt{T_w/\tau}$, yet their occurrence is  highly inhomogeneous.
Indeed, it follows from the arcsine law, see e.g. \cite{Feller}, that a time window of extent $\Delta t \ll T_w$
centered around a given whiff, typically contains $\sqrt{\Delta t/\tau}$ other whiffs. This number is much larger than $(\Delta t /T_w) \sqrt{T_w/\tau}$, which would hold
 for a homogeneous distribution. We conclude that short whiffs tend to cluster and to be interspersed by equally short blanks. Outside of the clusters, large excursions of the Lagrangian puffs generate long whiffs and blanks.
Whenever the probability of detecting a whiff is of order unity,  $T_\off \sim T_w$, there is symmetry between whiffs and blanks and individual clumps are virtually indiscernible.
Conversely, clumps stand out when the detection probability is small --  either because the point of detection lies outside of the average plume, or because the threshold of detection is large. Clumps are then sparsely distributed as a Poisson process with expected waiting time between clumps $\langle t_\off \rangle \simeq \langle t_w \rangle/\mathrm{Prob}(c>c_{\thr}) \gg \langle t_w \rangle$ (see \eqref{eq:Toff}), as expected from the Poisson clumping heuristics \cite{A89}.

\paragraph*{\bf Effects of the molecular diffusivity.--}
Differences in transport among various constituents of a blend are due to their 
molecular diffusivity $\kappa$. For small volatile compounds, such as pheromones, typical values for $\kappa$ are of the order $10^{-6}$ m$^2$/s, corresponding to P\'eclet numbers $UL/\kappa$ exceeding unity by several orders of magnitude \cite{CW08}.  Values of $\kappa$ do depend on the molecules, though, and their diffusion can thus be different. However, turbulent flows typically lead to the separation of Lagrangian particles (the exponent $\gamma$ is positive). Then, the effects of molecular diffusion are weak for large 
P\'eclet numbers and they are felt only at small separations among particles \cite{FGV01}.  The transition between the two regimes of transport occurs at the diffusive scale, which is in the range of a few millimeters to the centimeter (thus below the size of the source) for relevant flows \cite{CW08}. We conclude that the statistics of the concentration depends weakly on $\kappa$ and, most importantly, that the species-specific information on the ratios among constituents of a blend of molecules is largely preserved as the mixture is carried by turbulent flow. 
These conclusions are also supported by experimental data on laboratory flows, where the weak dependency on $\kappa$ of the concentration statistics was investigated and quantified \cite{D10}. 

\paragraph*{\bf Persistence of odor blends.--}
When female moths of different species emit blends composed of the same constituents but with different ratios, their messages may interfere and impair the correct decoding by male moths (see Fig.~1c). The goal of this Section is to clarify the conditions ensuring that interference does not occur. 

We consider a set of sources of size $a$, spaced by a distance $d \gg a$ from each other, emitting different blends of the same chemical compounds. Each 
source $k=1,2,\ldots$ releases the chemical species $i=1,2,\ldots$ at a rate $J^{(k)}_i$ (all rates are assumed comparable). 
The Lagrangian approach prescribes that we should follow the evolution of
a puff released at the detection point and traveling backwards in time. If the puff hits one and only one source, then the resulting signal can be
unambiguously attributed to it. Conversely, if the puff traverses two or more sources, the concentration is a mix of their emissions.
Given a detection threshold $c_\thr$, 
of the same order for all the components, the probability of receiving a mixed signal equals 
the probability that a puff 
crosses two sources while keeping the same (small) size.
The condition for a proper identification of the blend is derived in the Appendix A and reads\,:
 \begin{equation}
 \label{mainunmixcond}
 1\lesssim \left(\frac{R(d)}{r_{\thr}}\right)^{\gamma}=\left(\frac{c_{\thr}}{C(d)}\right)^{\frac{\gamma}{3-\alpha}}=\left(\frac{c_{\thr}}{C(x)}\right)^{\frac{\gamma}{3-\alpha}}\frac{d}{x}\,.
 \end{equation}

For typical concentrations $c_{\thr}\simeq C(x)$ and the probability of receiving a mixed signal
reduces to $p_{mix} \sim \exp[-(d/x)^\beta]$\, (see eq.~\eqref{pmix}): in order to discriminate two different sources by sampling typical concentrations, their separation $d$ must be comparable to the distance $x$ separating the receiver from one source. Our prediction agrees with experimental observations where the cross correlation between the concentration of two scalars emitted by different sources was measured \cite{KDV13}. Conversely, intense events carry more information
and allow to tell closer sources apart. 
Indeed, (\ref{mainunmixcond}) shows that whiffs with strong concentrations $c \gtrsim c_0 (d/a)^{-(3-\alpha)/\gamma}$  are unmixed -- they carry the proportion of constituents of only one source at any given time. Therefore,
the larger the threshold of detection, the greater the power of discrimination (at the expense of sensitivity and time) and vice versa. Even though we have not pursued detailed applications here, Lagrangian methods for the transport of blends can be relevant for the design of mating disruption for pests and disease-transmitting vectors 
\cite{codling, vectors}.

\medskip
\section{Results: Numerics and experiments}
\label{sec:appl}
To test our predictions, we  consider three different types of turbulent flows.

\paragraph*{\bf Kraichnan flow .--} This is a stochastic velocity field, incompressible, homogeneous and isotropic, with Gaussian statistics, uncorrelated in time, and self-similar Kolmogorov-Richardson spatial scaling (see \cite{FGV01} for review). These properties correspond to the exponents $\alpha=2$, $\beta=1$ and $\gamma=2/3$ defined in our formulation of the problem.
The advantage of this idealized model is that the Lagrangian Montecarlo method in \cite{FMV98} allows the numerical simulation of the integer moments of concentration for 
conditions (namely the ratio $a/x$ between the size of the source and the distance from it)  that are prohibitive for a fully-resolved integration of the fluid-dynamical equations. 
In summary, the results for the concentration statistics along the wind axis are (see the Appendix B for a detailed derivation)
\begin{equation}\label{summary:kraichnan}
C(x) \sim x^{-3/2}\;; \qquad \chi \sim \sqrt{\frac{x}{L}} \;\qquad \left\langle c^n \right\rangle \sim \chi C^n \sim x^{-(3n-1)/2}
\end{equation}
Fig.~3a shows that the first four moments are in excellent agreement with the theoretical prediction above.

\paragraph*{\bf Jet flow. --}
This is a laboratory flow qualitatively similar to wind tunnel experiments. Even though distances from the source are moderate compared to olfactory searches, experimental data still provide a compelling test for our general theory.  
For the experimental set-up in Ref.~\cite{VI99}, the single-particle motion is governed by large-scale components of the flow and $\alpha=1$. The main contribution 
to the dispersion of Lagrangian puffs arises from rapid, small-scale velocity fluctuations that induce a diffusive separation ($\gamma=2$) with diffusivity $k'\sim va$. Stationarity and homogeneity of the flow ensure $\beta=1$. The function $f$ in Eq.~\eqref{maineq:0}
is derived in the Appendix B. In addition to a crosswind Gaussian decay, it contains a prefactor $\left(U/v\right)^2$ which reflects the semi-conical shape of the average plume, with aperture angle $v/U$. The area of impact with the source is therefore amplified by $\left(U/v\right)^2$ with respect to an isotropic distribution.

The expressions just listed imply that along the mean axis
\begin{equation}\label{summary:jet}
 C(x) \sim J/(vax)\;; \qquad \chi \sim Ua/vx\;; \qquad  \langle c^n \rangle \sim \chi C^n\propto x^{-n-1}\;.
\end{equation} 
 with the rate $J \simeq c_0 U a^2$.  
The scaling of the moments of the concentration is in agreement with experimental data in Fig.~3b.
Fig.~3c presents the distribution of the concentration at various distances along the mean wind axis, compared to our 
 prediction  
\begin{equation}\label{summary:pcjet} 
p(c) \sim (\chi/c) \exp(-c/C)
\end{equation}
from \eqref{maineq:1}. Experimental data for the duration of whiffs and blanks are compared to Eqs.~\eqref{maineq:3} and \eqref{maineq:4} in Figs.~3d-e and f. 
The most likely duration $\tau$ is the time to diffuse across the source $\tau \sim a^2/k'\sim a/v$. We show in the Appendix B that the dominant mechanism that cuts off long whiffs is the 
large-scale sweeping of the Lagrangian puffs and we provide there the expression for the cutoff $T_w$ in the exponential function $g_w$ in Eq.~\eqref{maineq:3}.
Blanks obey the 
power-law predicted by \eqref{maineq:4} over nearly two decades. The Poisson clumping 
regime is realized at distances where detections are sparse and thus $\chi$ is small, i.e. $x \gg aU/v$ along the mean wind axis. In that regime, the exponential in $p(c)$ implies that the average duration of blanks depends  exponentially on the threshold $c_\thr \gtrsim C$\,: $\langle t_b\rangle  \sim \langle t_w \rangle \exp\left(c_\thr/C\right)$. Note that $\langle t_b\rangle$ grows exponentially with the distance to the source as well, since $C\propto 1/x$.

\paragraph*{\bf Atmospheric boundary layer --}
Finally, we consider the near-neutral atmospheric surface layer \cite{KF94}, the case most directly relevant for olfactory
searches. Two particular features of this flow are\,: (i) the mean wind depends logarithmically on the height $z$ above the ground; (ii) velocity fluctuations have their intensity $v$ nearly constant yet their correlation length is proportional to $z$. 
The consequence of (i) is that the time to transport particles from the source to the detection point is approximately $t_{\hit}\sim (x/U)\log^{-1}(z/h)$, where $h$ is the roughness height \cite{KF94}. The resulting modification to $t_{\hit}$ should a priori be applied to our formulae but in practice it is safely ignored as the logarithmic factor varies slowly. Consequences of (ii) are more conspicuous as the increase of the correlation length results in an effective diffusivity $\simeq vz$. Power counting gives then that $z$ is proportional to time and the growth of the effective diffusivity with $z$ implies the ballistic growth of both the single-particle displacement and the separation between pairs of particles, i.e. $\alpha=\gamma=1$.  The rate-of-growth of a puff of size $r$ is $\zeta_r \sim (v^2 t)/r^2$, 
which corresponds to $\beta=2$. These scalings are confirmed by experiments with puffs released in the atmospheric surface layer \cite{YKB98}.

Inserting the values above into \eqref{maineq:0}, we obtain  (see Appendix B)
\begin{equation}\label{summary:abl}
\chi \sim \cosh^{-2}\left( \frac{Uy}{vx} \right)\; \qquad C \sim c_0 \left(\frac{Ua}{vx}\right)^2
\end{equation}
that the intermittency factor $\chi$ is independent of the downwind distance $x$ and decays exponentially in the crosswind direction $y$, as confirmed in Figs. 4a-b. The figures show experimental data \cite{MM91,Y93} for the fluctuation intensity $\sigma_c/\langle c \rangle\sim \sqrt{\chi^{-1}-1}$, where $\sigma_c$ is the standard deviation of the concentration.
Eq.~\eqref{maineq:0} also predicts for the typical concentration in a whiff $C \sim c_0(Ua/vx)^2$, where we estimate again $J \simeq c_0 U a^2$. Unfortunately, measurements of absolute concentration are marred by calibration issues \cite{Y93} so that the prediction cannot be tested directly. However,  eq.~\eqref{maineq:1} predicts for the tail of the probability distribution $p(c) \sim \left(\chi/c\right) \exp(-c/C)$ and therefore the  detection probability 
\begin{equation}\label{summary:abl-cum}
\mathrm{Prob}(c > c_{\thr})\sim \chi \Gamma(0,c_{\thr}/C)
\end{equation} 
(where $\Gamma$ is the incomplete Gamma function). The latter quantity is reliably measured as it depends on ratios of concentration and is in agreement  with data in Fig.~4c from two independent field experiments \cite{MM91,Y93}.  

As for  dynamical aspects of the signal, atmospheric data \cite{YCK95}  in Fig. 4d present a clear power-law distribution of the duration of the whiffs, in agreement with \eqref{maineq:3}. The typical duration of the whiffs $\tau\sim a^2 U/(v^2 x)$ is predicted to be independent of the threshold. 

Comparing the two possible mechanisms for the cutoff of the whiffs (see Appendix B) we find that it is determined by the dispersion due to turbulent mixing \textcolor{black}{ $ T_w  \sim Cx/Uc_\thr$}. This prediction is in qualitative agreement with experimental data (see Fig. 6 in \cite{Y93}); a quantitative comparison would require more statistics as $T_w$ is dominated by low-probability events. Apparently, the statistics of blanks was not measured in field experiments. However, the distribution for the duration of upcrossing intervals $t_u$, i.e. the time elapsed between the beginning of two consecutive whiffs, is available from \cite{YCK95} (see Fig. 4e). 
Our theory predicts for $t_u$ the same distribution as for the time intervals
between odd (or even) zeros of a random walk, which is again a power law $t_u^{-3/2}$ for  $\tau\lesssim t_u \lesssim T_w$, in agreement with experimental data.

\smallskip
We conclude with a summary of the formulae for the atmospheric boundary layer
relevant for the final discussion\,:
 \begin{equation}
 \label{summary}
 x_{\thr}\simeq \frac{aU}{v}\sqrt{\frac{c_0}{c_{\thr}}}\,;\quad \tau\simeq \frac{a}{v}\sqrt{\frac{c_{\thr}}{c_0}}\,; \quad \langle t_w\rangle\simeq \frac{a}{v}\sqrt{\frac{c_0}{c_{\thr}}}\,.
 \end{equation}
 The first equation gives the largest distance $x_{\thr}$ where the two conditions 
$\chi\sim 1$ and $c_{\thr}\simeq C(x)$ are satisfied. The first condition is verified along the mean wind axis, while the crosswind decay of $\chi$ defines the width of the detection cone $vx/U$. The average duration $\langle t_b \rangle$ of the blanks is comparable to $\langle t_w\rangle$ 
inside the cone $y/x<v/U$, while $\langle t_b \rangle\gg \langle t_w \rangle$ 
outside.

\section*{Discussion and conclusions}

We first consider the implications of our results for the olfactory response of insects. The detection region -- 
 where the message sent by female to male moths is least garbled by the turbulence transporting the pheromones -- is defined by two conditions\,: (i) the whiffs of pheromones are sufficiently frequent (that is, the intermittency factor $\chi$ defined in \eqref{maineq:0} is not small); (ii) the typical concentration $C$ in a whiff is detectible, i.e. its ratio $C/c_\thr$ with respect to the detection threshold $c_{\thr}$ is not negligible. Experimental measurements show that
{\it B. mori} males respond to air streams 
containing as little as 200 molecules of bombykol per cm$^3$, corresponding to a sensitivity threshold $c_\thr \sim 10^{-18}$ M \cite{BKS65}.  Measured rates for the emission of pheromones by female moths are of the order of few picograms per second (see e.g.~\cite{B80}),
 which correspond to an emission rate $J \sim 10$ fmol/s for a molecular weight of a few hundreds Daltons, typical for most pheromones. 
The corresponding concentration at the source is $c_0\sim 1$ pM, for a mean wind $U\sim 1$ m/s and a size $a$  of the source of a few centimeters, as typical for female moths. 

The physiological parameters above can be inserted 
into the results for the atmospheric surface layer that we derived here and summarized in Eq.~\eqref{summary}. We find that the detection region is a semiconical volume (with aperture angle controlled by the ratio 
between the intensity of turbulent fluctuations and the mean wind) that extends to downwind 
distances $x_{\thr}\sim 10^3$ m, in  agreement with 
observations \cite{WP87}.  
 Hundreds of meters away from the source, the most likely duration $\tau$ of the whiffs is a few milliseconds, which compares well to the 
shortest pulses detectible by moths \cite{K13}. At those distances, whiffs tend to occur in clusters and a time window of $1$ second centered around a detection, typically contains $10-20$ odor encounters. This information-ridden pattern of stimulation is time-integrated at the level of the projection neurons and plays an important role in enhancing the behavioral sensitivity and in promoting exploitative sustained upwind flight \cite{K13,MC94}.  Upon approaching the source while 
staying inside the detection cone, the duration of the clumps decreases proportionally to the distance to the source. 
As a result, the search process is expected to lead to a statistically self-similar set of flight trajectories. 
Outside the detection cone, blank periods without any detection of pheromones are typically much longer than the whiffs. Note that even inside the detection cone, periods below threshold might be very long and last up to hundreds of seconds. 

When relatively long blanks occur,  
moths switch to cueless, exploratory casting phases, see e.g. \cite{BV97}. 
While surges are straightforward to define as upwind motion, the trajectories during casting phases are more involved. For example, the angle of flight with respect to the mean wind, the duration of crosswind extensions, their dependencies on the duration of the ongoing blank period, are all factors that potentially affect the patterns of flight during the casting phases. In particular, the extent of the memory of past detections that affect the casting  is an open issue. Another open issue is whether or not spatial information on the location of previous detections is involved in the control of the casting (and how, if positive).
Search strategies for olfactory robots \cite{VVS07} have shown that extended temporal memory and maps of space do lead to effective searches that alternate surges and casting phases qualitatively resembling those of insects. However, neurobiological constraints were not considered. The upshot is that quantitative data on flight patterns and their relation 
to the history of detections are needed to make progress on the decision-making processes controlling  the casting of insects during their olfactory searches. 

Laboratory bioassays with olfactometers and tethered insects are poised to shed light on  the previous issues by jointly analyzing the time history of odor stimuli and the virtual flight trajectories of the insects.
To ensure that responses observed in the laboratory are informative about the actual behavior of insects, it is crucial though that stimuli be as close as possible to those experienced by insects in the open field. This is the practical level where our results will be useful\,: 
Eqs.~\eqref{maineq:0}, \eqref{maineq:1},  \eqref{maineq:3} and \eqref{maineq:4} define a complete protocol for generating a sequence of odor pulses, whose intensity/duration and spacing should follow the statistics for whiffs and blanks, respectively.  The generation of such stimuli for  the neutral atmospheric boundary layer
seems achievable mechanically (see \cite{DCF09,B10}) or optogenetically \cite{K13} without stringent limitations.   Trains of odor stimuli having such distributions provide a statistically faithful representation of the landscape of odor detections created by atmospheric turbulence during olfactory searches. 
The protocol we derived here should then inform the design of future olfactory stimulation assays.

\section{Appendix A: Theory}

\setcounter{equation}{0}

\renewcommand{\theequation}{A\arabic{equation}}

\paragraph*{\bf The intensity of the concentration signal.--} 

We first consider the statistics of the pheromone concentration $c$ at a fixed time. 
If the puff hits the source, it overlaps with it for a time $t_s$, which depends on the size $r_{\hit}$ of the puff at the time $t_{\hit}$ of the hit. Since trajectories forming a puff are statistically equivalent and the flow is incompressible,  their weight is uniform.  Furthermore,
we neglect small-scale structures and we assume that the scaling of the volume of the puff is statistically determined by its size $r_{\hit}$ only, i.e. it is $\propto r_{\hit}^3$. Here and in the sequel we neglect constants of order unity. It follows that the concentration $c$ is the random function
\begin{equation}\label{eq:M3}
c \propto \left\{
\begin{array}{ll} 
J t_{s} r_{\hit}^{-3} & \mbox{with probability $p_{\hit}$,} \\ 
0 & \mbox{otherwise,}
\end{array}
\right.
\end{equation}
of the random variable $r_{\hit}$.
The expressions for the probability $p_{\hit}$ of hitting the source, $t_{\hit}$ and $t_s$ follow from the single-particle dispersion defined in the Section~\ref{expos}. 
For $r_{\hit} \ll x$, 
we have 
\begin{equation} \label{eq:M4}
t_{s} \sim \frac{r_{\hit}^\alpha}{k}\,; \qquad t_{\hit}\sim \frac{x}{U}\,;\qquad
p_{\hit} \equiv \left(\frac{r_{\hit}}{x}\right)^{3-\alpha}  f\left(\frac{U y^\alpha}{k x} \right)\,.
\end{equation}
The expression of $p_{\hit}$ above is justified by the fact that single-particle trajectories dispersing as $t^{1/\alpha}$ have (fractal) dimension $\alpha$  and the hitting probability  scales then with the codimension 
$3-\alpha$ appearing in $p_{\hit}$ \cite{Uriel}. For $\alpha=2$, $p_{\hit}$ in Eq.~\eqref{eq:M4} gives the well-known expression 
for the hitting probability of a random walk with drift, at large distances from the source \cite{Leal}. 
The function $f$ in (\ref{eq:M4}) is non-dimensional and decays rapidly as its argument becomes large, i.e. moving crosswind away from the wind axis. 

While the center of the puff is moving backward in time toward the source, its size grows as $r\sim (k't)^{1/\gamma}$. 
The size $r_{\hit}$ when the puff hits the source is expected to have a self-similar distribution \cite{Uriel,CGHKV03}, i.e. its expression reads 
\begin{equation}\label{eq:M5}
p(r_{\hit}) = \frac{1}{R(x)} \phi\left(\frac{r_{\hit}}{R(x)}\right)\,;\quad {\rm with}\quad R(x)\sim\left(\frac{k' x}{U}\right)^{1/\gamma}\,,
\end{equation}
denoting the typical size of the puff at $t_{\hit}$ defined by (\ref{eq:M4}).
The precise form of $\phi$ is unknown and depends on the details of the turbulent flow, yet its asymptotic behavior
is derived as follows. The probability that the size $r_{\hit}$ is well below its typical value ($r_{\hit}\le r\ll R(x)$) is
\begin{equation}\label{eq:M6}
\int_0^r p(r')dr' \sim e^{-\int_0^{t_{\hit}} \zeta_{r,t'}dt'} \sim e^{-(k' t_{\hit})^\beta r^{-\beta\gamma}}\,.
\end{equation}
The first step in eq.~(\ref{eq:M6}) states that the probability that a puff does not grow beyond the size $r$ is given by a Poisson process with local time rates $\zeta(t')= d\log r(t')/dt'$. The crucial physical ingredient justifying the use of Poisson statistics is that since $r\ll R(x)$, the total time $t_{\hit}$ is much longer than the typical time for growth at the scale $r$. Therefore, the total probability is the product of many largely independent factors. The second step in (\ref{eq:M6}) simply follows from the definition of $\beta$. 
Differentiating \eqref{eq:M6} with respect to $r$ and replacing $t$ by $t_{\hit}\sim x/U$, we finally obtain 
\begin{equation}\label{eq:M8}
p(r_{\hit}) \sim \left(\frac{k' x}{U r_{\hit}^\gamma}\right)^\beta \frac{1}{r_{\hit}}  \exp\left[-\left(\frac{k' x}{Ur_{\hit}^\gamma}\right)^\beta \right]\;\;\Rightarrow\; \phi(\rho) \sim \rho^{-1-\beta\gamma} \exp\left(-\rho^{-\beta\gamma}\right)\;\; {\rm for}\; \rho \ll 1\,.
\end{equation}

Eqs.~\eqref{eq:M3} and \eqref{eq:M4} imply that the mean concentration $\langle c \rangle=J t_s r_{\hit}^{-3} p_{\hit}$ 
does not depend on $r_{\hit}$, which reflects the conservation of mass. Conversely, averaging $J t_s r_{\hit}^{-3}$ with respect to \eqref{eq:M5} (which amounts to replacing $r_{\hit}$ by $R(x)$, apart from numerical factors)
gives the conditional average concentration 
\begin{equation}\label{eq:00}
 C(x)=\left\langle c | c>0 \right\rangle \sim \frac{J}{k}R(x)^{\alpha-3}\,\, \sim 
 \frac{J}{k} \left(\frac{k'x}{U}\right)^{-\frac{3-\alpha}{\gamma}}\,.
\end{equation}
Finally, averaging $p_{\hit}$ in Eq.~\eqref{eq:M4} over $r_{\hit}$, we obtain for the intermittency factor 
$\chi$ 
\begin{equation}\label{eq:0}
\chi =\mathrm{Prob}(c>0)\sim 
\left(\frac{k' x^{1-\gamma}}{U}\right)^{\frac{3-\alpha}{\gamma}}f\left(\frac{U y^\alpha}{k x} \right)=\left(\frac{R(x)}{x}\right)^{3-\alpha}f\left(\frac{U y^\alpha}{k x} \right)\,.
\end{equation}
Eqs.~(\ref{eq:00}) and (\ref{eq:0}) show that, as $y$ increases, $\chi$ decreases and $C$ remains constant.
Therefore, moving crosswind away from the mean-wind axis, the signal retains its intensity but becomes sparser.
Approaching the source (reducing $x$), the intensity within a whiff grows ($\alpha<3$), 
while the frequency of encounters depends on $\gamma$. 

It follows from \eqref{eq:M3} that the probability distribution of the concentration $p(c)$ 
contains two terms. The first one is the singular contribution at the origin $\delta(c)\int\left[1-p_{\hit}\right]p(r_{hit})\,dr_{\hit}$. The second one is the continuous contribution $p_{\hit}p(r_{hit})|dr_{\hit}/dc|$. The relation between $r_{\hit}$ and $c$ is read from the first line in \eqref{eq:M3} and more conveniently recast as 
\begin{equation}
\label{rhitc}
r_{\hit}=R(x)\left(\frac{c}{C(x)}\right)^{-\frac{1}{3-\alpha}}\,.
\end{equation}
By using (\ref{eq:M4}), (\ref{eq:M5}) and (\ref{eq:0}), we finally obtain
\begin{equation}\label{eq:M11}
p(c) = (1-\chi)\delta(c) +  \chi p^+(c)\quad {\rm with}\quad p^+(c)= \frac{1}{(3-\alpha)C} \left(\frac{c}{C}\right)^{-2-\frac{1}{(3-\alpha)}} \frac{\phi\left( \left(\frac{c}{C}\right)^{-\frac{1}{(3-\alpha)}} \right)}{\int\phi(u)u^{3-\alpha}\,du}\,.
\end{equation}
Intense concentrations are associated to flow configurations which leave the  puff atypically small (see (\ref{eq:M3}) for $\alpha<3$). Since those rare configurations obey the Poisson asymptotics \eqref{eq:M8}, the tail of the probability distribution $p(c)$ is\,:
\begin{equation}\label{eq:1}
p(c) \sim \frac{\chi}{C} \left(\frac{c}{C}\right)^{-2+\frac{\beta\gamma}{3-\alpha}} \exp\left[-\left(\frac{c}{C}\right)^{\frac{\beta\gamma}{3-\alpha}}\right]\quad {\rm for} \quad C \ll c \ll c_0  \,,
\end{equation}
where $c_0$ is the concentration at the source. 

Finally, it follows from Eq.~\eqref{eq:M11} that the moments $\langle c^n\rangle$ depend on $C$ and $\chi$ in \eqref{eq:00} and (\ref{eq:0}) as \begin{equation}
\langle c^n \rangle \sim \chi C^n\propto x^{-\left(3-\alpha\right)\left(1+\left(n-1\right)/\gamma\right)}\,.
\label{cn}
\end{equation}

\paragraph*{\bf The duration of the whiffs.--}
\label{durwhiffs}
As discussed in the Introduction, the behavior of insects depends on the time-course of the odor stimuli. It is therefore important to characterize the statistics of the whiffs, i.e. time intervals when the concentration is above the threshold $c_{\thr}$ of detection. The complementary intervals when $c\leq c_{\thr}$ are dubbed ``blanks'', or ``below threshold'' since the signal is either absent or not detectible. The ratio $C/c_{\thr}$, with $C$ given by (\ref{eq:00}), determines whether a typical plume is detectible. 

We consider a time $t^*$ when the concentration $c({\bm x},t)$ just exceeded the threshold $c_{\thr}$. 
We are interested in the statistics of the duration $t_w$ of the whiff, that is the time interval such that the concentration stays above the threshold for its whole duration and falls below at $t^*+t_w$.
The single-particle exponent is taken $\alpha=1$, since the laboratory and the atmospheric flow
 that we shall analyze below have that value (see Sections~\ref{sec:jet} and \ref{sec:ABL}). Our prediction 
for the distribution of  the duration of the whiffs reads
\begin{equation} \label{eq:3}
p(t_w) \sim  \frac{1}{\tau} \left( \frac{t_w}{\tau} \right)^{-3/2} g_w\left(t_w\right)\,.
\end{equation} 
The function $g_w$ is constant for small arguments and decays rapidly for $ t_w> T_w$, where the cutoff $T_w$ is determined below.  The decay law of $g_w$ is exponential with rate $T_w^{-1}$ for $ t_w \gtrsim T_w$.  Note that, due to the slow power-law decay $-3/2$ in  \eqref{eq:3}, \textcolor{black}{the average duration is determined by the cutoff\,: $\langle t_w \rangle \sim (\tau T_w)^{1/2}$. }

\smallskip
Let us now derive (\ref{eq:3}). From (\ref{eq:M3}), (\ref{eq:M4}) and (\ref{rhitc}) for $\alpha=1$, a threshold $c_{\thr} \gtrsim C$ is associated to a size of the puff
\begin{equation}
\label{rthr}
r_{\thr}\sim\sqrt{\frac{J}{k c_{\thr}}}=\sqrt{\frac{C(x)}{c_{\thr}}}R(x)\,.
\end{equation}
As time progresses, the turbulent velocity field ${\bm v}$, as well as the probability $p_{\bm v}$ in \eqref{eq:M2}, evolve. For two times spaced by $t_w$, the two puffs to be tracked are released from the same position but with a delay $t_w$. The delay decorrelates the trajectories of the two puffs 
as we proceed to quantify. It is convenient to discuss separately the effects of the scales of the turbulent flow larger, comparable or smaller than $r_{\thr}$. 

(i) Large-scale (sizes $\gg r_{\thr}$)
velocity fluctuations transport the puffs almost uniformly and their major effect is then to displace the puffs. The differential displacement between the puffs released at different times can lead to termination of the whiff by making the later puff (released at $t^*+t_w$) lose overlap with the source (see Fig.~2d). The typical time for the loss of contact is determined by analyzing the dynamics of lateral displacements. 
The angular size of the puff as seen from the detection point is $r_{\thr}/x$. The trajectories of particles transported by the flow form angles (with respect to the direction of the mean wind) of typical amplitude $v/U$. The angle fluctuates in time with a correlation frequency $v/L$, where $L$ is the correlation length of the flow. The rate of change of the angle is thus $v^2/LU$. Finally, we combine the two terms above 
and insert the expression (\ref{rthr}) of $r_{\thr}$.
We conclude that 
 the typical time for a lateral displacement of the puffs leading to a loss of contact with the source is  
 \begin{equation}
 T_{\textit{displace}}  \sim \frac{UL}{v^2}\frac{r_{\thr}}{x}\simeq \frac{UL}{v^2}\frac{R(x)}{x}
 \sqrt{\frac{C(x)}{c_{\thr}}}\,.
 \label{T2}
 \end{equation}

(ii) Scales comparable to the size of the puff $r_{\thr}$ are less effective at displacing puffs yet they
disperse them, i.e. enlarge their size. That effect can terminate the whiff by making 
$r_{\hit} > r_{\thr}$ for the later puff released at $t^*+t_w$ (see Fig.~2d). The characteristic time for the growth of the size of a puff is\,:
\begin{equation}
T_{\textit{disperse}}= \zeta_{r_{\thr},t_{\hit}}^{-1}  = \frac{x}{U}\left(\frac{R(x)}{r_{\thr}}\right)^{-\beta\gamma}=\frac{x}{U} \left(\frac{C(x)}{c_{\thr}}\right)^{\frac{\beta\gamma}{2}}\,,
\label{T1}
\end{equation} 
where we used the definition of $\beta$, $t_{\hit}\simeq x/U$ and (\ref{eq:M5}) for the second equality and (\ref{rthr}) for the last. For times $t_w\gg T_{\textit{disperse}}$, we can treat successive time intervals of length $T_{\textit{disperse}}$ as largely independent, use again Poisson statistics (as for \eqref{eq:M6}) and obtain that the probability 
for the size to remain below $r_{\thr}$ 
for the whole interval $(t^*,t^*+t_w)$ is $e^{-\int_0^{t_w}\zeta_{r_{\thr},t_{\hit}}\,dt'}\sim e^{-t_w/T_{\textit{disperse}}}$. A similar  reasoning can be used for (i) and yields an exponential decay as well. 
 
Both  physical mechanisms (i) and (ii) lead to a smaller cutoff as the threshold $c_{\thr}$ is increased. Their relative strength depends on the turbulent flow transporting the odors, on the distance to the source and on the threshold $c_{\thr}$, as discussed in the examples below. The cutoff $T_w$ in \eqref{eq:3} is the minimum between $T_{\textit{displace}}$ and $T_{\textit{disperse}}$.

\smallskip
(iii) Small scales play a crucial for the dynamics of two puffs delayed by times $t_w<T_w$. We are interested in situations where $c_{\thr}\gg C(x)$, i.e. intense fluctuations only are detectible. In the Lagrangian formulation, those rare events correspond to puffs reaching $r_{\thr}$, defined in (\ref{rthr}), and then maintaining that size for an anomalously long time. Indeed, the typical time to reach the size $r_{\thr}$ is $r_{thr}^\gamma/k'$. The time to reach the source is $x/U$.  Using (\ref{eq:M5}) and (\ref{rthr}),  their ratio is  $(C(x)/c_\thr)^{\gamma/2} << 1$, i.e. most of the time to reach the source is spent with sizes $\sim r_{\thr}$. 

The characteristic time $\zeta_{r_{\thr},t}\ll t_{\hit}$ for any time $t<t_{\hit}$. The total displacement of the puff at $t_{\hit}$ is thus the sum of largely uncorrelated events of typical amplitude $r_{\thr}$ and analogous to a diffusion process with effective diffusivity 
\begin{equation}
D_{\thr}\sim r^2_{\thr} \frac{\int_0^{t_{\hit}} \zeta_{r_{\thr},t'}\,dt'}{t_{\hit}}\sim r^2_{\thr}\, \zeta_{r_{\thr},t_{\hit}}\,.
\label{Dthr}
\end{equation}

We now consider two puffs released with an initial time delay $t_w$. The time delay induces an initial spatial separation $\sim Ut_w$. For the cases we shall consider, it can be verified that even for the smallest times $\tau$ in (\ref{eq:3}), the initial separation $U\tau$ is larger than the viscous scale of the flow. Therefore, velocity fluctuations smaller than the size of the two puffs are uncorrelated since the very beginning of the trajectories. At the time when the size $r_{\thr}$ is reached, the displacement of the two trajectories is $\gtrsim r_{\thr}$ and subsequent displacements are then largely independent.  We conclude that at the time $t_{\hit}$ when the puffs reach the source, their centers ${\bm x}_c$ are separated as for a three-dimensional diffusion process with the diffusivity $\sim D_{\thr}$ in (\ref{Dthr}) (a factor two is neglected, as all other constants of order unity). 

The beginning of a whiff occurs when the entire source of size $a$ is barely within the puff released at $t^*$, i.e. the center of the source is at a distance $\sim a$ from the boundary of the puff. The end of the whiff occurs when the center of the source first loses overlap with the puff. The centers ${\bm x}_c$ of the puffs released at times later than $t^*$ 
displace diffusively with coefficient $D_\thr$, as shown above. We conclude that $t_w$ is distributed as the first exit time for a diffusing process \cite{Feller}, which obeys the $-3/2$ power law in \eqref{eq:3}.
The shortest time $\tau$ in \eqref{eq:3} corresponds to the fastest exit\,:
%
\begin{equation}
\tau\simeq \frac{a^2}{D_{\thr}}= \zeta_{r_{\thr},t_{\hit}}^{-1} \left(\frac{a}{r_{\thr}}\right)^2\propto x^{1-2/\gamma} \left(\frac{c_{\thr}}{C}\right)^{1-\beta\gamma/2}\,.
\label{tau}
\end{equation}
 The time to diffuse across the whole size of the puff is $r_{\thr}^2/D_{\thr}=\zeta_\thr$, coinciding with the typical time for the dispersion of the puff to larger sizes.
For $c_\thr \lesssim C$ the size of the puff saturates to its typical value $R(x)$ (see \eqref{eq:M5}), with no dependency on the threshold $c_{\thr}$.

\paragraph*{\bf The duration of periods below threshold.--} Blanks are time intervals when the concentration is below $c_\thr$ and thus no signal is detectible. Our prediction for the probability density of their duration $t_\off$ is
\begin{equation}\label{eq:4}
p(t_\off) \sim \frac{1}{\tau} \left(\frac{ t_\off}{\tau} \right)^{-3/2} g_\off(t_\off)\,.
\end{equation}
Here, $g_\off$ is constant for $t_b$ smaller than the cutoff $T_\off$ and then decays exponentially with  rate $T_\off^{-1}$.
The physical origin of the $-3/2$ power law in \eqref{eq:4} is identical to \eqref{eq:3}, i.e. the diffusion on rapid time scales of the Lagrangian puff, that symmetrically loses and gains contact with the source. Remark  that the power laws do not depend on the details of the flow. The temporal structure of whiffs and blanks contains then some information independent of environmental variations of the intensity, stratification and other details of the flow transporting the pheromones. 

It follows from \eqref{eq:4} that \textcolor{black}{ the average duration $\langle t_\off \rangle\sim (\tau T_\off)^{1/2}$, i.e. it is determined by the cutoff of the distribution, as for the whiffs.}
Whiffs and blanks are mutually exclusive so that their averages (and their cutoffs) are not independent. In particular, the probability of detection equals the average fraction of time spent above the threshold $c_{\thr}$\,:
\textcolor{black}{
\begin{equation}
\label{eq:Toff}
\mathrm{Prob}(c>c_\thr)=\frac{\langle t_w \rangle}{\langle t_w \rangle+\langle t_\off \rangle}\quad\Rightarrow\quad (\tau T_\off)^{1/2}  \sim \langle t_\off \rangle = \langle t_w \rangle \frac{\mathrm{Prob}(c\le c_\thr)}{\mathrm{Prob}(c>c_\thr)} \sim (\tau T_w)^{1/2} \frac{\mathrm{Prob}(c\le c_\thr)}{\mathrm{Prob}(c>c_\thr)} 
\end{equation}
}
Eq.~\eqref{eq:00} shows that $C$ (and the statistics of $t_w$, see \eqref{eq:3}) does not change with the crosswind distance, i.e. intensity and duration of the whiffs are independent of $y$. Their frequency changes, though, as shown by the intermittency factor $\chi$ \eqref{eq:0} and the statistics of the intervals below threshold. Namely, \eqref{eq:Toff} indicates that the cutoff $T_{\off}$  grows moving crosswind. 

\paragraph*{\bf Persistence of odor blends.--}
When female moths of different species emit blends composed of the same constituents but with different ratios, their messages may interfere and impair the correct decoding by male moths (see Fig.~1c). The goal of this Section is to clarify the conditions ensuring that interference does not occur. 

We consider a set of sources of size $a$, spaced by a distance $d \gg a$ from each other, emitting different blends of the same chemical compounds. Each 
source $k=1,2,\ldots$ releases the chemical species $i=1,2,\ldots$ at a rate $J^{(k)}_i$ (all rates are assumed comparable). 
Eq.~\eqref{eq:M2} states that we should follow the evolution of
a puff released at the detection point and traveling backwards in time. If the puff hits one and only one source, then the resulting signal can be
unambiguously attributed to it. Conversely, if the puff traverses two or more sources, the concentration is a mix of their emissions.
Given a detection threshold $c_\thr$, of the same order for all the components, the probability of receiving a mixed signal equals 
the probability that a puff of the size $r_\thr$ given by \eqref{rthr} crosses two sources while keeping the same size.
Clearly, if $r_\thr \gtrsim d$, mixing of the signals is almost certain. When $r_\thr \lesssim  d$, the probability of  mixing 
is the product of the probability that the puff is not dispersed, multiplied by the probability for a particle starting from one source to hit the other. The worst case scenario is when the various sources are aligned along the mean wind. The probability of a mixed signal is then the product of $p_\hit$ in (\ref{eq:M4}) and (\ref{eq:M6}) (with $x$ replaced by $d$, $y=0$ and $r=r_\thr$)\,:
\begin{equation}
p_{\rm mix}\simeq \left(\frac{r_\thr}{d}\right)^{3-\alpha}\,\times \exp\left[-\left(\frac{k'd}{Ur_\thr^{\gamma}}\right)^\beta\right]\,.
\label{pmix}
\end{equation}
The mixing probability is
 small if $r_\thr \ll d$ or $r_\thr\ll \left(k'd/U\right)^{1/\gamma}$. The right-hand side in the last inequality is recognized by (\ref{eq:M5}) as  the typical separation $R(d)$ between particles in the time $d/U$ to travel the distance $d$ between the sources. Typically, $R(d)\ll d$ for $\gamma\ge 1$ and the condition for a proper identification of the blend is then\,:
 \begin{equation}
 \label{unmixcond}
 1\lesssim \left(\frac{R(d)}{r_{\thr}}\right)^{\gamma}=\left(\frac{c_{\thr}}{C(d)}\right)^{\frac{\gamma}{3-\alpha}}=\left(\frac{c_{\thr}}{C(x)}\right)^{\frac{\gamma}{3-\alpha}}\frac{d}{x}\,,
 \end{equation}
having used the relation \eqref{eq:00} between size of the puff and concentration. 

For typical concentrations, $c_{\thr}\simeq C(x)$ and the exponential factor in (\ref{pmix})
reduces to $\exp[-(d/x)^\beta]$\,: in order to discriminate two different sources by sampling typical concentrations, their separation $d$ must be comparable to the distance $x$ separating the receiver from one source. Our prediction agrees with experimental observations where the cross correlation between the concentration of two scalars emitted by different sources was measured \cite{KDV13}. Conversely, intense events carry more information
and allow to tell closer sources apart. 
Indeed, (\ref{unmixcond}) shows that whiffs with strong concentrations $c \gtrsim c_0 (d/a)^{-(3-\alpha)/\gamma}$  are unmixed -- they carry the proportion of constituents of only one source at any given time. Therefore,
the larger the threshold of detection, the greater the power of discrimination (at the expense of sensitivity and time) and vice versa. Even though we have not pursued detailed applications here, Lagrangian methods for the transport of blends can be relevant for the design of mating disruption for pests and disease-transmitting vectors 
\cite{codling, vectors}.

\section{Appendix B: Numerics and experiments}

\setcounter{equation}{0}

\renewcommand{\theequation}{B\arabic{equation}}

To test our predictions, we  have considered three different types of turbulent flows (Kraichnan flow, jet flow and neutral atmospheric boundary layer) that we proceed to discuss. 

\paragraph*{\bf Kraichnan flow.--}

Kraichnan flow (see \cite{FGV01} for review) are stochastic velocity fields, incompressible, homogeneous and isotropic, with Gaussian statistics, uncorrelated in time, and self-similar Kolmogorov-Richardson spatial scaling. 
The advantage of this idealized model is that the Lagrangian Montecarlo method in \cite{FMV98} allows the numerical simulation of the integer moments of concentration for ratios $a/x$ (the size of the source over the distance to it)  that are prohibitive for a fully-resolved integration of the fluid-dynamical equations. 
 
 The (unrealistic) short time-correlation of the Kraichnan flow induces the diffusion of single particles 
\cite{FGV01}. The corresponding diffusivity is $k\sim L^{4/3}\sim v L$, where $L$ is the correlation length of the flow and $v$ is the typical amplitude of the Gaussian velocity fluctuations. At short distances, diffusion dominates over the mean wind $U$, which takes over  at distances $\sim L\,v/U$. Parameters are chosen to ensure $x\gg L\,v/U$ so that the time to reach the source is still $t_{\hit}\simeq x/U$. 
The single-particle exponents defined in Section~\ref{expos} are then $\alpha=2$ and $k\sim vL$. The spatial scaling of the velocity differences ensures that pair dispersion obeys the Richardson-Kolmogorov scaling $\gamma=2/3$; the constant $k' \sim v/L^{1/3}$. This scaling behavior holds as long as the separation among the particles remains below $L$ (diffusion sets in at larger separations).
Finally, homogeneity and stationarity ensure $\beta=1$. 

Inserting the values above into  \eqref{eq:00}, \eqref{eq:0} and (\ref{cn}), we obtain 
\begin{equation}
\label{Kra}
C(x) \propto x^{-3/2}\,;\quad \chi \propto \sqrt{\frac{x}{L}}\,;\quad \langle c^n \rangle \sim \chi C^n \propto x^{-(3n-1)/2}\,.
\end{equation}
The cutoff function $f$ in Eq.~\eqref{eq:0} is obtained from results on diffusive processes (see, e.g., \cite{Leal} and Supplementary Material (SM)) as $f(\xi)=\exp(-\xi/4)$. The scaling of the moments (\ref{Kra}) holds when the typical size of the puffs at $t_{\hit}$ is larger than the size of the source $a$. Otherwise, all the moments $\langle c^n \rangle$ tend to coincide with the probability $p_{\hit}=a/x$ that a diffusing particle hit a sphere of size $a$, starting at distance $x$ from it. The predictions (\ref{Kra}) for the scaling of the first four moments shown in Fig.~3a are in excellent agreement with the results of numerical simulations.

The numerical method used to simulate the moments $\langle c^n \rangle$ relies on taking the $n$-th power of 
\eqref{eq:M2} and averaging over the velocity field to obtain the moments in terms of Lagrangian trajectories. Using the short correlation of the Kraichnan flow,
 the trajectories of $n$ particles generated by a Montecarlo method are sufficient to obtain the $n$-th order moment \cite{FMV98}. We used a numerical implementation identical to \cite{FMV98}, referred to for details. 

The new difficulty is that most trajectories miss the source and a small fraction of the statistical realizations contribute. Indeed, \eqref{eq:M2} shows that realizations where at least one of the $n$ particles misses the source do not contribute to the moments. 
We circumvented the problem by using importance sampling \cite{MD01}.  Namely, we chose one reference particle and sampled its trajectories by generating a Brownian bridge (see \cite{BB} and SM) between the starting point and the source. This guarantees that the particle hit the source at least once. The time of the first passage at the source is generated from the exact probability distribution, calculated by standard methods (see SM for details).  The remaining $n-1$ particles evolve according to the exact dynamics for 
their relative separations, as in \cite{FMV98}. 

The general idea of importance sampling \cite{MD01} is that the quality of a Montecarlo estimation improves if the auxiliary 
distribution is more concentrated on the subset of events which substantially contribute to the observable being measured. In our case, the number of statistical samples required for the Montecarlo estimation of $\langle c^n \rangle$ is reduced by the factor $p_{\hit}$ defined in (\ref{eq:M5}).
For the simulations in Fig. 3a, the gain is of the order $x/a\simeq 10^4-10^5$ along the wind axis and it further increases with the crosswind distance. Further details on the method can be found in SM.

\paragraph*{\bf Jet flow.--}
\label{sec:jet}

We now consider experimental data for a jet flow \cite{VI99}, a setup qualitatively similar to wind tunnel experiments. Even though distances from the source are moderate compared to those for olfactory searches by moths, experimental data still provide a compelling test for our general theory.  
The experimental flow \cite{VI99} is well modeled by the superposition of a mean flow $U\simeq 0.8$ m/s and a statistically homogeneous, isotropic flow with correlation length $L\simeq 8$ cm and intensity of the fluctuations $v/U\simeq 0.25$.  The turbulence level is relatively high, which a priori affects our estimates, e.g. of the time and the probability of hitting the source. Nevertheless, we show below that our predictions agree with experimental data, suggesting that corrections mainly affect constants of order unity, which we have disregarded. 

Large-scale fluctuations of the flow decorrelate on a time scale $\sim L/v$. At distances $x \lesssim UL/v$, the time to reach the source $t_{\hit}\sim x/U<L/v$ and large-scale diffusion (with diffusivity $\simeq vL$), which could potentially dominate the transport of single particles, has not set in yet. For the experiments 
in \cite{VI99}, $UL/v$ is about $30$cm, which is $2-3$ times bigger than the largest distance to the source where measurements are made. It follows that large-scale fluctuations induce a ballistic motion of amplitude $v$ in each realization of the flow (for the relevant times $\lesssim t_{\hit}$). Since $v\ll U$, the ballistic contribution by  the mean velocity $U$ is stronger. 
Small-scale fluctuations have shorter correlations and do produce a diffusive motion. However, their diffusivity is $\sim va$ and the resulting displacement  
$\sqrt{va t_{\hit}}$ is negligible compared to $vt_{\hit}$ for $x\gtrsim Ua/v$. Since $a$ is a few millimeters, we conclude that single-particle parameters defined in Section~\ref{expos} are $\alpha=1$ and $k=U$. 
The main contribution 
to the dispersion of Lagrangian puffs stems from rapid, small-scale velocity fluctuations that induce a diffusive separation ($\gamma=2$) with diffusivity $k'\sim va$. The diffusive contribution $\sqrt{vat_{\hit}}$
 dominates Richardson's dispersion $\sqrt{v^3t^3_{\hit}/L}$ for $x \lesssim (U/v)\sqrt{aL}$. The latter takes over at larger distances. We conclude that for $a \lesssim vx/U\ \lesssim \sqrt{La}$, 
the size of the puff grows diffusively, i.e. $\gamma=2$, $k'=va$. Finally, stationarity and homogeneity of the flow ensure $\beta=1$. 

\smallskip
In summary, the parameters defined in Section~\ref{expos} are\,: $\alpha=1$, $k=U$, 
$\gamma=2$, $k'=va$ and $\beta=1$.
Inserting them into \eqref{eq:00} gives for the conditional average concentration $C$\,:
\begin{equation}
C \sim \frac{J}{vax}\,,\quad {\rm with}\quad J \simeq c_0 U a^2\,.
\label{Cjet}
\end{equation}

The function $f$ in Eq.~\eqref{eq:0}
is derived as follows. The probability $p_{\hit}$ of hitting the source is the probability that a spherical puff of size $r_{\hit}$, starting at $(x,y,0)$ (with $x \gg r_\hit$) and moving with constant velocity $(-U+v_x,v_y,v_z)$, 
hit the origin. The constancy of the velocity stems from the ballistic motion discussed above.
For a given ${\bm v}$, hitting occurs if at the time $t_\hit\simeq  x/U$ the distance of the center of the puff from the source is smaller than its radius\,: 
$\sqrt{(y+v_y t_\hit)^2+v^2_z t^2_\hit} < r_\hit$. 
The probability of satisfying this inequality in the space $(v_y,v_z)$ is calculated for a Gaussian, isotropic distribution of the fluctuations with standard deviation 
$v \ll U$. Using that the angle formed by the directions of the mean wind and the starting point of the puff is small (since $v/U$ is supposed small), we obtain\,:
\begin{equation}
\label{phit}
p_{hit} \simeq \left(\frac{Ur_{\hit}}{vx}\right)^2  e^{-  \left(\frac{Uy}{vx}\right)^2}\,;\quad
\chi \sim \frac{Ua}{vx} e^{-\left(\frac{U y}{vx}\right)^2}\,,
\end{equation}  
where we omitted constant factors. Along the wind axis $y=0$, $p_{hit}$ reduces to the ratio between the cross sectional area of the puff $\simeq r_{\hit}^2$ and the area $(vt_\hit)^2$, transverse to the wind axis, spanned by 
the center of the puff  at $t_\hit$. Comparing (\ref{phit}) to (\ref{eq:M4}), we identify the prefactor $\left(U/v\right)^2$ for the function $f$, which reflects the semi-conical shape of the average plume with aperture angle $v/U$. The area of impact with the source is thus amplified by $\left(U/v\right)^2$ with respect to an isotropic distribution. The second equation in (\ref{phit}) is obtained using (\ref{eq:0}), the expression of $f$ just discussed. 
 
The expressions (\ref{Cjet}) and (\ref{phit}) can be inserted into \eqref{eq:1} and \eqref{cn} to obtain\,:
\begin{equation}
p(c) \sim \frac{\chi}{c} \exp\left(-\frac{c}{C}\right)\,;\quad \langle c^n \rangle \simeq \chi C^n \sim c_0^n \left(\frac{Ua}{vx}\right)^{n+1}\,,
\label{pcjet}
\end{equation}
where the second expression is specified for the axis $y=0$. The resulting scaling behavior with respect to $x$ is in excellent agreement with experimental data in Fig.~3b. The prediction for the probability distribution of the concentration at various distances along the wind axis is also supported by experimental data shown in Fig.~3c. 

As for dynamical aspects of the signal of odors, the two cutoffs discussed in the section \ref{durwhiffs} read
\begin{equation}
T_{\textit{disperse}}\sim \frac{ac_0}{vc_{\thr}}\,;\quad T_{\textit{displace}}\sim \frac{aUL}{xv^2}\sqrt{\frac{c_{0}}{c_{\thr}}}\,.
\end{equation}
The latter is shorter than the former for sufficiently small thresholds and the cutoff in Eq.~\eqref{eq:3} is then $T_w=T_{\textit{displace}}$.
The cutoff $T_b$ in Eq.~\eqref{eq:4} for the duration of the blanks follows from the general relation \eqref{eq:Toff}.
The shortest duration 
$\tau$ in (\ref{tau}) 
is $\tau\sim a/v$, independent of the detection threshold and of the distance to the source. 

Experimental data for the duration of whiffs and blanks are compared to Eqs.~\eqref{eq:3} and \eqref{eq:4} in Figs.~3d-e and f. 
Blanks obey the predicted 
power-law $-3/2$ over nearly two decades. The Poisson clumping 
regime is realized where detections are sparse and thus $\chi$ is small, i.e. $x \gg aU/v$ along the mean wind axis. In that regime, the exponential form \eqref{pcjet} of $p(c)$ and the relation \eqref{eq:Toff} imply that the average duration of the blanks depends  exponentially on the threshold $c_\thr \gtrsim C(x)$\,: $\langle t_b\rangle  \sim \langle t_w \rangle \exp\left(c_\thr/C(x)\right)$. Note that $\langle t_b\rangle$ grows exponentially with $x$, since $C(x)\propto 1/x$.

\paragraph*{\bf Near-neutral boundary layer.--}
\label{sec:ABL}

We finally consider the near-neutral atmospheric surface layer \cite{KF94}. This is the case most directly relevant for olfactory
searches by moths as they usually search at dusk when convective effects are weak and no stable stratification is present. The latter is more typical at night while strong convective effects might be present during daytime. Stratification conditions generally depend on micro-meteorological conditions such as cloud coverage and humidity. We do not address these aspects here. It is worth noticing that some properties of the odor landscapes such as the shape of probability density function of whiff and blank durations turn out to be largely insensitive to such details.

\smallskip
Flow in the neutral boundary layer have two special features with respect to the previous cases\,: 

(i) The mean wind depends logarithmically on the height $z$ above the ground, {\it viz.}
$U(z)= (v/\varkappa) \log [(z - h)/z_0] $ 
where $v$ is the friction velocity, $\varkappa \simeq 0.4$ is the von Karman's constant, $z_0$ is the roughness height, $h$ is the displacement height (roughly two thirds of the canopy height) \cite{KF94}. Typical values in the atmospheric surface layer are $v \sim 0.1 - 0.5$ m/s, $z_0, h \sim 0.1 - 1$ m depending on whether land surface is covered by high grass, pastures or forests.
The consequence of the logarithmic profile is that the time to transport particles from the source to the detection point has a logarithmic dependency on the height. In practice, the resulting modification is safely ignored as the logarithmic factor varies slowly. We shall also omit the order unity von Karman constant. 

(ii) The intensity $v$ of velocity fluctuations  is nearly constant yet the size of the largest eddies at height $z$ is  $\propto z$ and their correlation time is $z/v$. We are interested in situations where particles are released close to the ground (heights much smaller than the Monin scale of the boundary layer \cite{KF94}). The variance of the displacement in the height $z$ behaves then as $d z^2/dt \sim v z $, i.e. ballistically due to the effective diffusivity $\simeq vz$. The average height will also systematically increase ballistically. Note that this last statement is due to particles being released close to the ground (the growth of the mean height saturates as its value becomes comparable to the Monin scale). Height and time (since the release of the particles) will therefore grow in parallel. Since the effective diffusivity behaves as $v z$, fluctuations in the lateral and longitudinal displacements grow proportionally to $t$ as well.   Along the wind direction, the mean wind $U\gg v$ dominates the transport and the hitting time is $t_\hit \sim x/U$. Similarly, the sweeping time of a puff of size $r_{\hit}$ across the source is $t_s \sim r_\hit/U$.  
The separation between a pair of particles is similarly determined by a height-dependent diffusion process with coefficient $vz$ and therefore scales as $vt$, which gives
the exponent $\gamma=1$ and $k'=v$.  The rate-of-growth for a puff of size $r$ is $ v z / r^2 \sim v^2 t/ r^{2}$, yielding
$\beta=2$. 

\smallskip
In summary, the parameters defined in the section \ref{expos} are\,: $\alpha=1$, $k=U$, 
$\gamma=1$, $k'=v$ and $\beta=2$.
These scalings are confirmed by experiments with puffs released in the atmospheric surface layer \cite{YKB98}.

It follows from Eq.~\eqref{eq:0} and the values above that the typical conditional concentration is 
\begin{equation}
\label{cABL}
C(x) \sim c_0 \left(\frac{Ua}{vx}\right)^2\,;\quad J\simeq c_0Ua^2\,.
\end{equation}
 For the intermittency factor $\chi$ in Eq.~\eqref{eq:0}, we need the form of $f$. 
The advection-diffusion equation with mean wind $U$ and diffusion coefficient $vz$ is solved in the SM by an eigenfunction expansion, as in the simpler case of constant diffusivity. The analytical solution confirms the scalings justified above intuitively and gives
\begin{equation}
\label{phitL}
p_{\hit} \simeq \left(\frac{Ur_\hit}{vx}\right)^2\frac{1}{\cosh^2\left(\frac{U y}{v x}\right)}\,; \quad \chi \sim \frac{1}{\cosh^{2}\left(\frac{Uy}{vx}\right)}\,.
\end{equation}
The probability $p_{\hit}$ decays exponentially in the crosswind direction $y$, determining the semi-conical shape of the average plume, with aperture angle $v/U$. 
The second equation in \eqref{phitL} is obtained by using (\ref{eq:0}). The intermittency factor $\chi$ is thus independent of  $x$ and decays exponentially in $y$, as confirmed in Figs. 4a-b.
The data from experiments in \cite{MM91,Y93} report the fluctuation intensity $\sigma_c/\langle c\rangle$, where $\sigma_c$ is the standard deviation of the concentration. Using (\ref{cn}) for the  moments of the concentration, $\sigma_c/\langle c\rangle\simeq \sqrt{\chi^{-1}-1}$. It follows from \eqref{phitL} that in the average plume, $y\lesssim vx/U$,  the fluctuation intensity is order unity, while outside the cone it grows exponentially, as observed in the data. Skewness and kurtosis also grow exponentially with the transverse distance  (in agreement with Figs. 2-4 in \cite{YCK95}). 


The tail of the probability density of the concentration follows from \eqref{eq:M11}
 and the scaling function $\phi(\rho)$ in \eqref{eq:M8}. 
 Upon insertion of the appropriate exponents $\alpha=\gamma=1$ and $\beta=2$, we obtain
\begin{equation}
\label{pcABL}
p(c) \simeq \frac{\chi}{c} e^{-c/C}\,\quad {\rm for} \quad c \gtrsim C.
\end{equation}
Unfortunately, measurements of absolute concentration are marred by calibration issues \cite{Y93} so that the prediction (\ref{pcABL}) cannot be tested directly. However, integration of $p(c)$ gives $\mathrm{Prob}(c>c_\thr)\sim \chi \Gamma(0,c_\thr/C)$, where 
$\Gamma$ is the incomplete Gamma function. The latter quantity is reliably measured as it depends on ratios of concentration and is in agreement  with data in Fig.~4c from two independent field experiments \cite{MM91,Y93}.  

\smallskip
As for  dynamical aspects of the signal, atmospheric data \cite{YCK95} in Fig. 4d display a clear $-3/2$ power-law for the duration of the whiffs, in agreement with \eqref{eq:3}. Using the general expressions (\ref{rthr}), (\ref{eq:M5}) and (\ref{tau}), we derive 
\begin{equation}
\label{tauABL}
\zeta_{r,t_{\hit}}\sim \frac{x}{U}\left(\frac{v}{r}\right)^2 \,;\quad r_{\thr} \simeq \frac{vx}{U} \sqrt{\frac{C(x)}{c_{\thr}}}\simeq a \sqrt{\frac{c_0}{c_{\thr}}}\,;\quad \tau \sim \frac{a^2 U}{v^2 x}\,.
\end{equation}
 Comparing the two mechanisms for the cutoff of the whiffs (see Section~\ref{durwhiffs}), we find that 
 \begin{equation}
 T_{\textit{disperse}}\simeq \zeta^{-1}_{r_{\thr},t_{\hit}}\simeq \frac{r_{\thr}^2U}{v^2 x} \,;\quad T_{\textit{displace}}\simeq \frac{r_{thr}}{v}\,.
 \label{compareTT}
 \end{equation}
 For the second equality, we used $UL\simeq vx$, since the integral scale $L$ is proportional to the height $z$, and $z/x\simeq v/U$.
 The dispersive time is shorter than the displacement time as long as $r_\thr \lesssim vx/U$, so that  \textcolor{black}{
 $T_w\sim T_{\textit{disperse}}$ }.  The cut-off linearly increases with $x$ when the threshold is kept proportional to the typical value $C(x)$ (itself $\propto 1/x^2$), with a prefactor 
inversely proportional to the relative threshold. Our prediction is in qualitative agreement with experimental data (see Fig. 6 in \cite{Y93}); a quantitative comparison would require more statistics as $T_w$ is dominated by low-probability events. Apparently, the statistics of periods below threshold was not measured in field experiments. However, the distribution for the duration of upcrossing intervals $t_u$, i.e. the time elapsed between the beginning of two consecutive whiffs, is available from \cite{YCK95}. 
Our theory predicts for $t_u$ the same distribution as for the time intervals
between odd (or even) zeros of a random walk, which is again a power law $t_u^{-3/2}$ for  $\tau\lesssim t_u \lesssim T_w$, in agreement with experimental data in Fig. 4e. The average duration $T_w$ of the blanks follows from (\ref{compareTT}) and the relation \eqref{eq:Toff}.

\smallskip
We conclude with the derivation of the formulae relevant for the 
 final discussion\,:
 \begin{equation}
 \label{summary-app}
 x_{\thr}\simeq \frac{aU}{v}\sqrt{\frac{c_0}{c_{\thr}}}\,;\quad \tau\simeq \frac{a}{v}\sqrt{\frac{c_{\thr}}{c_0}}\,; \quad  \textcolor{black}{T_{\it disperse}\simeq \frac{a}{v}\sqrt{\frac{c_0}{c_{\thr}}}\,.}
 \end{equation}
 The first equation gives the largest distance $x_{\thr}$ where the two conditions 
$\chi\sim 1$ and $c_{\thr}\simeq C(x)$ are satisfied. It follows from \eqref{phitL} that the first condition is verified along the mean wind axis, while the crosswind decay of $\chi$ defines the width of the detection cone $vx/U$. Equating (\ref{cABL}) to $c_{\thr}$, we obtain $x_{\thr}$ in \eqref{summary}. The second and third equations in \eqref{summary} are the expressions (\ref{tauABL}) of $\tau$  and (\ref{compareTT}) of $T_{\textit{disperse}}$, estimated at $x=x_{\thr}$.
Finally, it follows from \eqref{eq:Toff} that the average duration $\langle t_b \rangle$ of the blanks is comparable to $\langle t_w\rangle$ 
inside the cone $y/x<v/U$, while $\langle t_b \rangle\gg \langle t_w \rangle$ 
outside.

\begin{acknowledgments}
We thank  P. Lucas, J.B. Masson, D. Rochat, J.P. Rospars for useful discussions. 
This work was funded by the French state program ÒInvestissements d'Avenir,Ó managed by the Agence Nationale de la Recherche (Grant ANR-10-BINF-05).
\end{acknowledgments}

\newpage

\begin{figure}
\includegraphics[width=9.5cm]{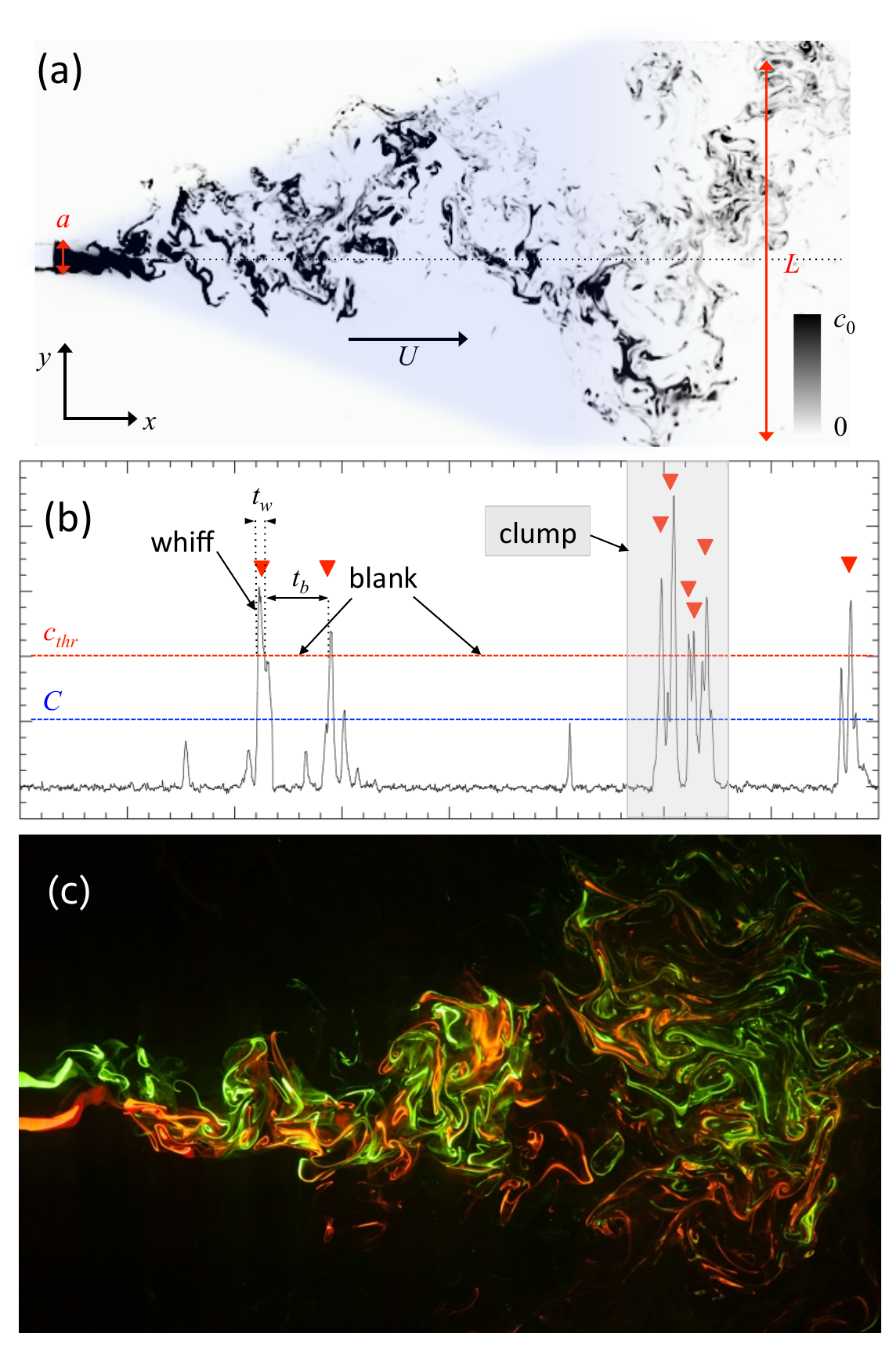}
\caption{The structure of a turbulent odor plume. (a) A two-dimensional section of a plume from the jet flow experiment \cite{VI99}. The shaded area is the projection of the conical average plume, i.e. the region outside of which crosswind transport is weak and the odor concentration decays rapidly. (b) A typical time-series of the odor concentration at a given point in space \cite{VI99}. Red triangles indicate the occurrence of whiffs, i.e. intervals when the local concentration is above the threshold $c_{\thr}$ indicated by the red line. For  olfactory searches the threshold is comparable to the sensitivity of the pheromone receptors of the insects. The blue line indicates the average concentration $C$ in the regions where the signal is above the noise level. (c) A two dimensional section of two blending plumes from the jet flow experiment \cite{KDV13}. The two different chemicals mix as they progress downwind and the 
resulting signal is a blend.
}
\end{figure}
\begin{figure}[!b]
\includegraphics[width=9.5cm]{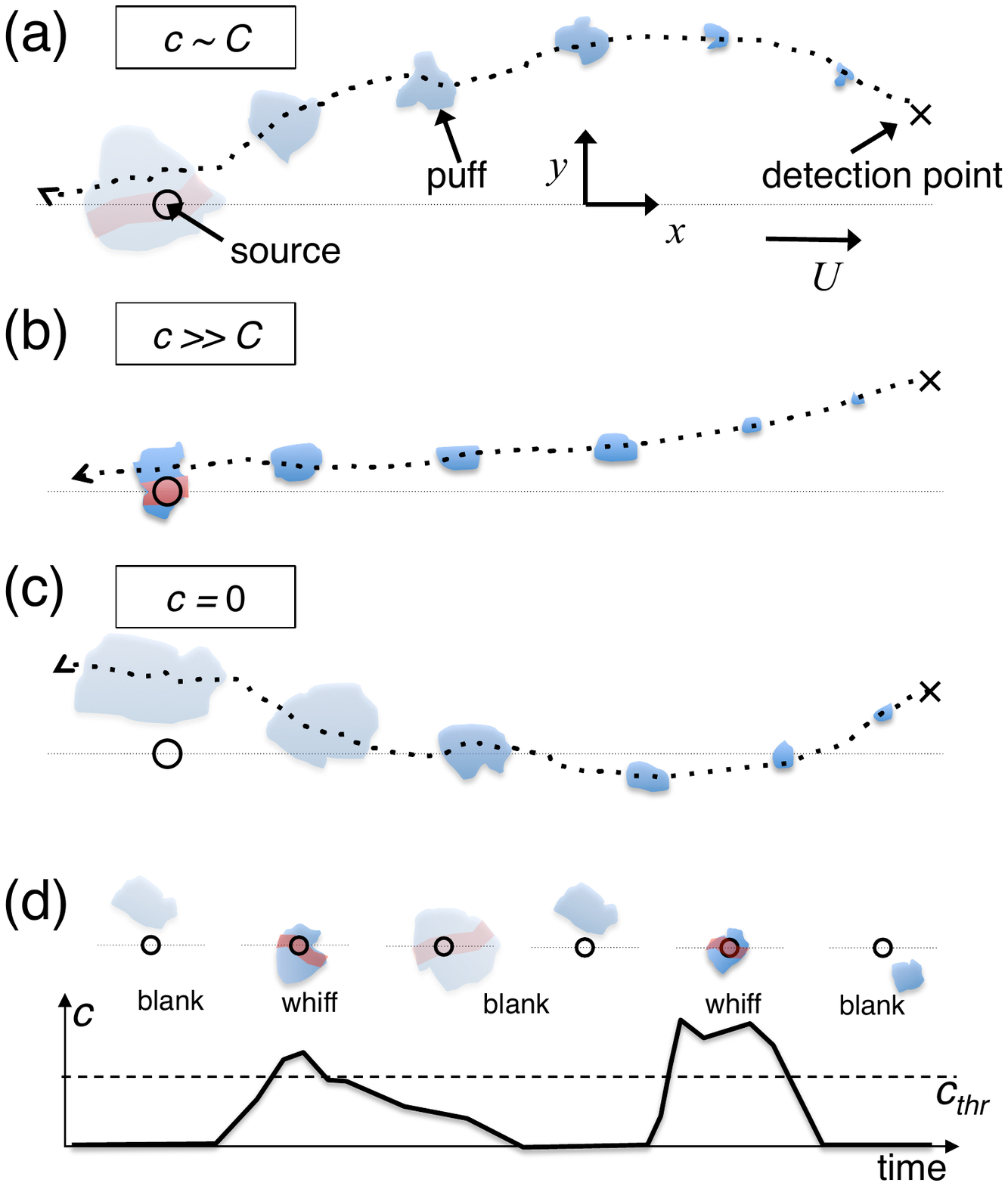}
\caption{Scheme of the Lagrangian approach. The concentration $c$ at a given location ${\bm x}$ and time $t$ is expressed in terms of the history of a Lagrangian puff, that is, an ensemble of particles transported by the turbulent flow, all starting at ${\bm x}$ at time $t$ and dispersing backwards in time. The concentration $c$ is determined by the size of the Lagrangian puff when it hits the source (if it does)\,: (a) Average values $c\simeq C$ correspond to the puff hitting the source with a typical value of the size; (b) Intense concentrations $c$ correspond to the puff hitting the source with unusually small sizes; (c) The concentration $c$ vanishes if the puff never hits the source throughout its history. (d) The sketch of a time series. From left to right: blank: the concentration $c$ vanishes; whiff: the puff hits the source with a small size and $c$ passes the threshold of detection $c_{\thr}$; blank:
turbulent diffusion enlarges the size of the puff and $c$ decays below the threshold, then $c$ vanishes due to the puff loosing contact with the source. The red strips indicate the regions of the puff overlapping with the source  as the puff is swept by the turbulent flow.
}
\end{figure}
\begin{figure}[!t]
\includegraphics[width=11cm]{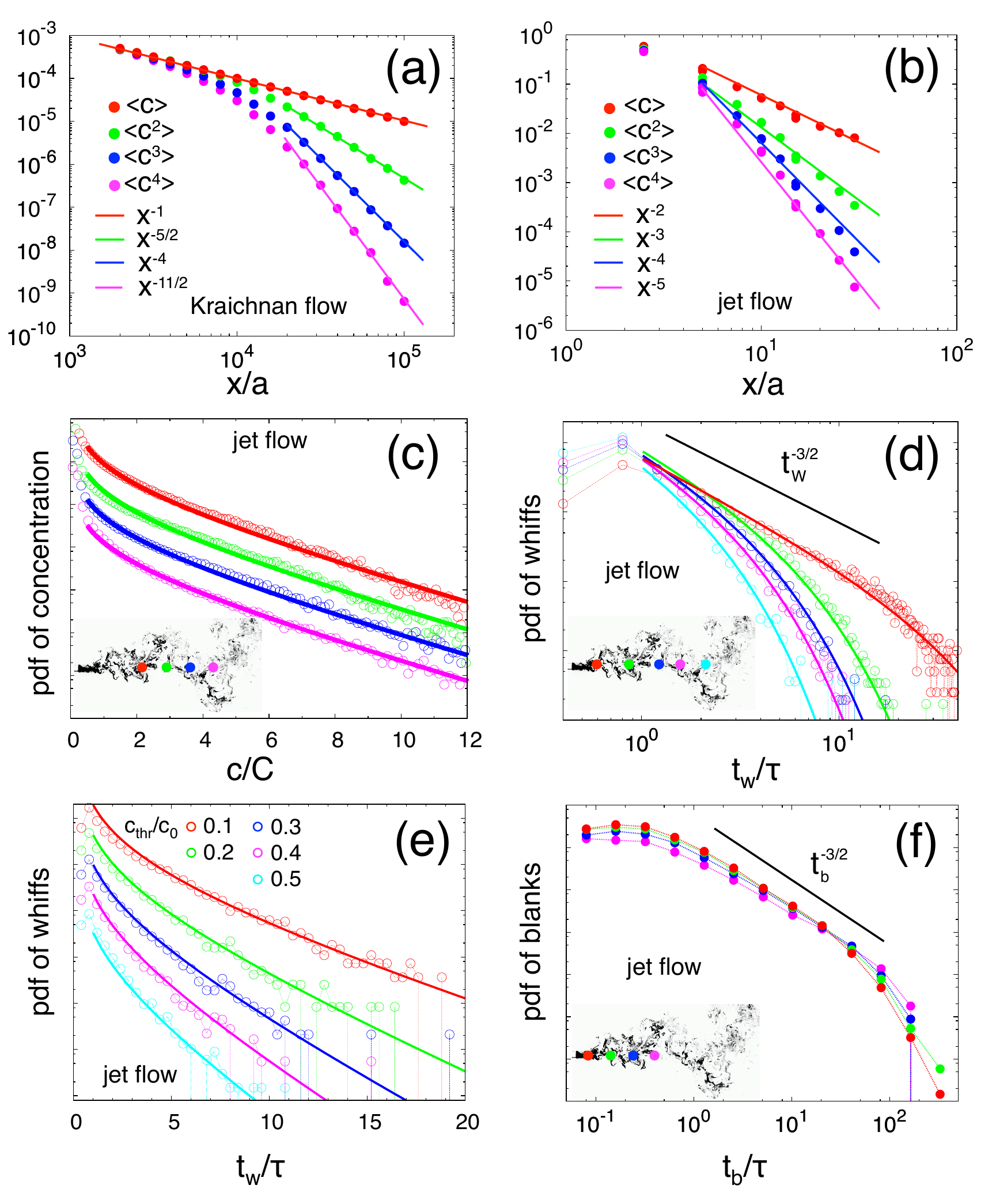}
\caption{The statistics of the concentration of odors for the Kraichnan flow \cite{FGV01} (panel a) and jet flow experiments \cite{VI99} (all other panels). (a) Moments of the concentration for the Kraichnan flow, as a function of the ratio $x/a$ between the distance $x$ to the source and its linear size $a$.
Full lines are the theoretical predictions in \eqref{summary:kraichnan}. (b) As in (a), for the jet flow experimental data, compared to our predictions \eqref{summary:jet}. (c) The Probability Density Function (pdf) of the concentration (rescaled by its typical value $C$ within the whiffs)
at various distances (shown in the inset) from the source, compared to our prediction \eqref{summary:pcjet}. Data have been shifted vertically for viewing purposes. 
(d) Pdf for the duration $t_w$ of the whiffs (time intervals when the concentration remains above a threshold of detection $c_{\thr}$) at various distances from the source. Full lines are our predictions \eqref{maineq:3}. Durations are rescaled by their most likely value $\tau$.  (e) The pdf for the duration of the whiffs {\it vs} $c_{\thr}$ (at  $x/a=5$), compared to our prediction \eqref{maineq:3}. (f) The pdf for the duration of the blanks (intervals without detections) at various distances from the source (shown in the inset), compared to our prediction \eqref{maineq:4}.}
\end{figure} 
\begin{figure}[!b]
\includegraphics[width=12cm]{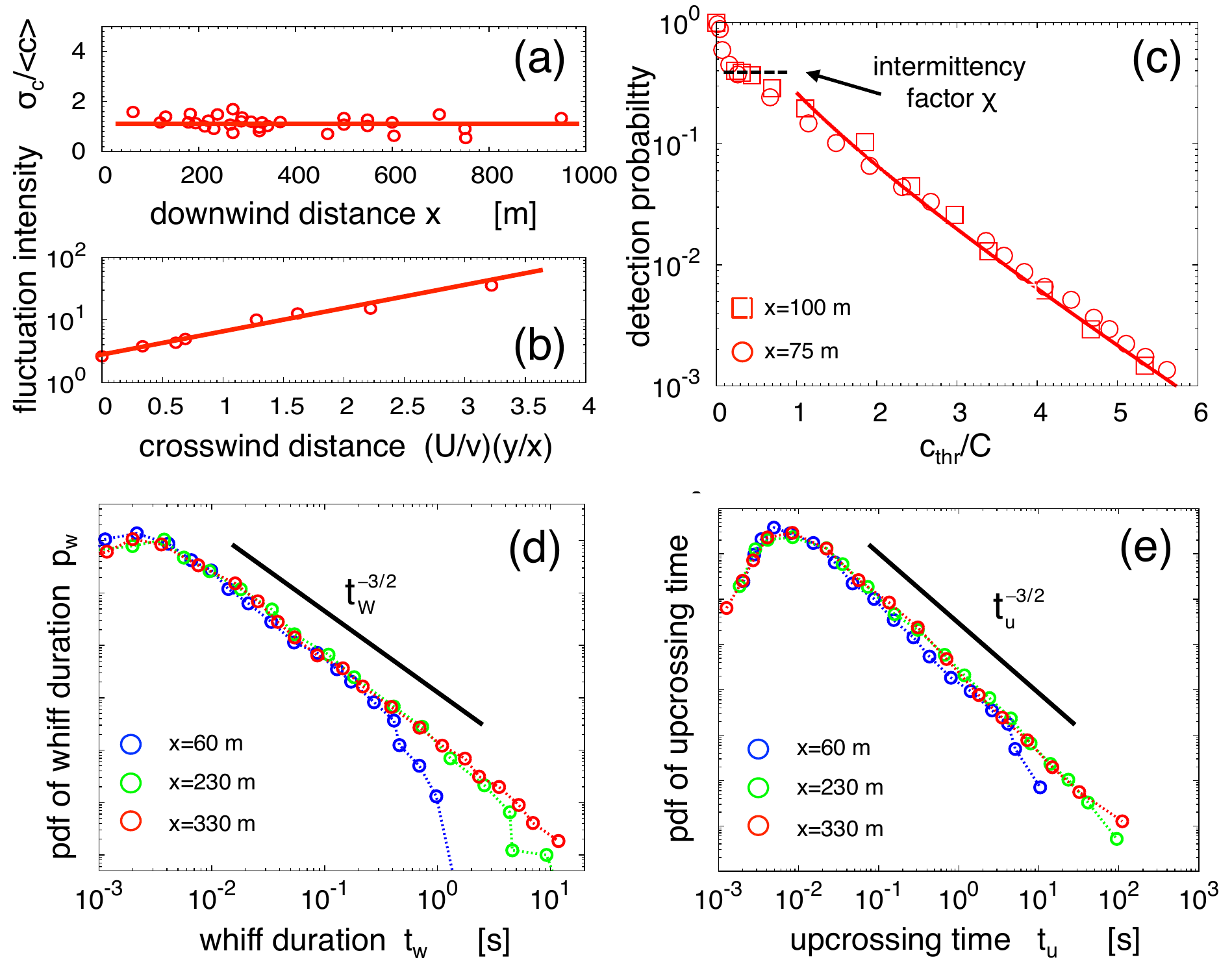}
\caption{Odor statistics in the atmospheric surface layer. (a-b) The intensity of the fluctuations of odor concentration as a function of the downwind distance $x$ and of the ratio between the crosswind distance $y$ and $x$, multiplied by the ratio between the mean $U$ and the turbulent component $v$ of the flow transporting the odors.
Solid lines are our predictions, from eq.~\eqref{summary:abl}. (c) The probability of detection, i.e. that the local concentration of odors is above a certain threshold, {\it vs} the value of the detection threshold. The full line is the theoretical prediction from~\eqref{summary:abl-cum}. The dashed line is the intermittency factor $\chi=\mathrm{Prob}(c>0)$ in \eqref{summary:abl}.
(d-e) The probability density functions for the duration of the whiffs (time intervals when the concentration remains above the threshold of detection) and  the upcrossing time intervals, defined as the time elapsed between the beginnings of two successive whiffs ($t_w+t_b$ in Figure 1b). The power-law $-3/2$ is the theoretical prediction derived here.}
\end{figure}

\end{document}